\documentclass[aps,prb,reprint,superscriptaddress]{revtex4-2}
\pdfoutput=1
\usepackage{easy-todo}
\usepackage{color}
\usepackage{graphicx}
\usepackage{amsmath}
\usepackage{amsfonts}
\usepackage{amssymb}
\usepackage{microtype}
\usepackage{verbatim}
\usepackage{bbold}
\usepackage{hyperref}
\usepackage{cleveref}
\allowdisplaybreaks
\usepackage{multirow}
\usepackage{bbm}
\usepackage{braket}
\hypersetup{
  colorlinks   = true, 
  urlcolor     = blue, 
  linkcolor    = blue, 
  citecolor   = blue 
}

\usepackage{textcomp}

\usepackage[smallerops]{newtxmath}

\newcommand*{\cL}{{\mathcal L}}
\newcommand*{\cC}{{\mathcal C}}
\newcommand*{\s}{\mathsf{S}}
\newcommand*{\g}{\mathsf{g}}
\newcommand*{\q}{\mathsf{q}}
\newcommand*{\Q}{\mathsf{Q}}

\begin{document}

\author{Pieter W. Claeys}
\email{pc652@cam.ac.uk}
\affiliation{TCM Group, Cavendish Laboratory, University of Cambridge, Cambridge CB3 0HE, UK}

\author{Austen Lamacraft}
\affiliation{TCM Group, Cavendish Laboratory, University of Cambridge, Cambridge CB3 0HE, UK}

\title{Dissipative dynamics in open XXZ Richardson-Gaudin models}

\begin{abstract}
In specific open systems with collective dissipation the Liouvillian can be mapped to a non-Hermitian Hamiltonian. We here consider such a system where the Liouvillian is mapped to an XXZ Richardson-Gaudin integrable model and detail its exact Bethe ansatz solution. 
While no longer Hermitian, the Hamiltonian is pseudo-Hermitian/PT-symmetric, and as the strength of the coupling to the environment is increased the spectrum in a fixed symmetry sector changes from a broken pseudo-Hermitian phase with complex conjugate eigenvalues to a pseudo-Hermitian phase with real eigenvalues, passing through a series of exceptional points and associated dissipative quantum phase transitions.
The homogeneous limit supports a nontrivial steady state, and away from this limit this state gives rise to a slow logarithmic growth of the decay rate (spectral gap) with system size.
Using the exact solution, it is furthermore shown how at large coupling strengths the ratio of the imaginary to the real part of the eigenvalues becomes approximately quantized in the remaining symmetry sectors.
\end{abstract}

\maketitle
\section{Introduction}

No quantum system is truly isolated, and in the past decade there has been an increasing interest in the physics of open quantum many-body systems. 
The coupling to an external bath can give rise to gain and loss terms and decoherence, which can no longer be described by Hermitian dynamics, and open systems are generally described by a quantum master equation in terms of a non-Hermitian Liouvillian \cite{breuer_theory_2007}.
However, dissipation is not necessarily detrimental -- the interplay between gain and loss can give rise to physics not accessible in closed systems and Hermitian dynamics. As one example, the system can exhibit PT symmetry when the gain and loss terms are exactly balanced \cite{mostafazadeh_pseudo-hermitian_2010,el-ganainy_non-hermitian_2018,huber_emergence_2020,nakanishi_pt_2021}, and this symmetry can be spontaneously broken, leading to exceptional points and dissipative quantum phase transitions \cite{heiss_physics_2012,ashida_non-hermitian_2020,bergholtz_exceptional_2021}.

While it is generally impossible to exactly solve the master equation, exact solutions for specific integrable Liouvillians have started to appear in the literature. These range from non-interacting systems \cite{prosen_third_2008,prosen_exact_2010,van_caspel_dynamical_2019,krapivsky_free_2019,shibata_dissipative_2019,vernier_mixing_2020,lieu_tenfold_2020,alba_noninteracting_2021} to boundary-driven systems \cite{prosen_open_2011,prosen_exact_2011,karevski_exact_2013,ilievski_dissipation-driven_2017,vanicat_integrable_2018,landi_non-equilibrium_2021} and systems that can be mapped to a non-Hermitian Yang-Baxter integrable Hamiltonian \cite{medvedyeva_exact_2016,shibata_dissipative_2019,ziolkowska_yang-baxter_2020,buca_bethe_2020,de_leeuw_constructing_2021}. A special class of the latter is those with collective dissipation \cite{rowlands_noisy_2018,ribeiro_integrable_2019,lerma-hernandez_trigonometric_2020,rubio-garcia_integrability_2021}, which will be the focus of this work. 
Specifically, we consider a Liouvillian that can be mapped to a Richardson-Gaudin integrable XXZ model. The integrability of this model was established by Rubio-Garc\'ia \emph{et al.} \cite{rubio-garcia_integrability_2021}, as a direct extension of earlier results of Ref.~\cite{rowlands_noisy_2018}, and subsequently used to study the spectral statistics as an indicator of quantum chaos in open systems. The Hermitian Richardson-Gaudin XXZ model is known to exhibit a rich phase diagram \cite{ortiz_exactly-solvable_2005,ibanez_exactly_2009,rombouts_quantum_2010,van_raemdonck_exact_2014,links_exact_2015,claeys_read-green_2016}, motivating a detailed study of the properties of the exact solution in the non-Hermitian case.

In this work, we detail the mapping to a non-Hermitian Hamiltonian and the subsequent exact solution of the model. This Hamiltonian, as well as the conserved charges, are shown to be pseudo-Hermitian, constraining the eigenvalues to be either real or part of a complex conjugate pair \cite{mostafazadeh_pseudo-hermiticity_2002,mostafazadeh_pseudo-hermitian_2010}. We show how eigenvalues change from complex conjugate pairs, when weakly coupled to the environment, to purely real eigenvalues when strongly coupled, after passing through exceptional points. At these exceptional points two eigenstates coalesce and the Liouvillian is no longer diagonalizable. Physically, these exceptional points correspond to dynamical dissipative phase transitions with non-analytic relaxation rates throughout the spectrum, including for the leading decay mode.

The homogeneous model admits a nontrivial steady state, and away from this limit we use the exact Bethe ansatz solution to show that in the inhomogeneous model the corresponding state decays with a small decay rate that scales logarithmically with system size. 
Here the exact Bethe ansatz solution is crucial in allowing us to obtain exact eigenvalues at system sizes out of reach of exact methods and establish this scaling. Somewhat surprisingly, in the purely dissipative regime  close to the exceptional point the decay can be slower than in the weakly coupled regime.
We additionally uncover a remarkable structure in the eigenspectrum, where at large coupling strengths the eigenvalues organize themselves on straight lines in the complex plane. The ratio of the imaginary to the real part of the eigenvalues becomes approximately quantized, indicating that the decay rate is proportional to oscillation frequency with a quantized proportionality factor.

This paper is organized as follows. In Section~\ref{sec:model} we introduce the model and its exact Bethe ansatz solution through the mapping to a non-Hermitian integrable model. Section~\ref{sec:spectrum} discusses the spectrum of the model and its implications on the dynamics, with special attention paid to its symmetry properties and the homogeneous and strong-coupling limits, after which the commuting quantities associated with integrability are discussed in Section~\ref{sec:charges}, as well as the numerical solution method for the Bethe equations. Section~\ref{sec:conc} is then reserved for conclusions.

\section{Model}
\label{sec:model}
Assuming Markovian bath dynamics, the Lindblad equation \cite{breuer_theory_2007} determines the time evolution of the density matrix $\rho$ as
\begin{equation}\label{eq:QME}
  \dot\rho = -i\left[H,\rho\right] + \sum_\alpha \left[L_\alpha \rho L^\dagger_\alpha-\frac{1}{2}L^\dagger_\alpha L_\alpha \rho-\frac{1}{2}\rho L^\dagger_\alpha L_\alpha  \right],
\end{equation}
where we set $\hbar = 1$ and choose the system Hamiltonian $H$ and the jump operators $L_{\alpha}, \alpha \in \{+,-,z\}$ as
\begin{align}\label{eq:model}
  &H = \sum_{j} \left[\Omega+\omega_j\right] s_j^z, \nonumber\\
  &L_z = \sqrt{\g_{0}}\sum_j s_j^z,\quad L_{\pm}= \sqrt{\g}\sum_j \sqrt{\omega_j} s_j^{\pm}\,.
\end{align}
The Hamiltonian describes a system of non-interacting spins, which we choose to have spin $1/2$, each subject to a local magnetic field. The $\pm$ terms describe collective gain and loss, and the $z$ term describes a collective dephasing. The amplitudes $\omega_j$ in the Lindblad operators are related to the amplitudes in the Hamiltonian in order to obtain a solvable model, which can physically be achieved by introducing a detuning of the magnetic fields proportional to the terms in the jump operators. While the Hamiltonian itself is non-interacting, the collective Lindblad operators lead to an interacting Liouvillian. Furthermore, the gain and loss are balanced with equal strength $\g$, whereas the dephasing is tuned by an independent prefactor $\g_0$.

As outlined in Ref.~\cite{rowlands_noisy_2018}, the Lindblad operator can be mapped to a non-Hermitian Hamiltonian acting on a doubled Hilbert space. E.g. a one-spin density matrix for spin $j$ can always be expanded as
\begin{equation}
\rho_j = \frac{1}{2}\mathbbm{1}_j + \sum_{\alpha} c_j^\alpha s^{\alpha}_j,
\end{equation}
with $c_j^{\alpha}$ complex coefficients and $s_j^{\alpha}$ spin-1/2 operators, and the operators can be mapped to spin singlet and triplet states as
\begin{equation}
\frac{1}{2}\mathbbm{1}_j \to \ket{0,0}_j,\quad s_j^z \to \ket{1,0}_j,\quad s^{\pm}_j \to \sqrt{2}\ket{1,\pm 1}_j,
\end{equation}
such that each spin operator maps to a spin-1 state. The Liouvillian acts trivially on the singlet states, guaranteeing that the identity is always a trivial steady state, and under this mapping the action of Eq.~\eqref{eq:model} on the triplet states can be described by an operator \footnote{Technically, this operator generates the evolution of the corresponding correlation functions, and its Hermitian conjugate generates the evolution of the operators. We write the Liouvillian in this way in order to make the connection with the literature.}
\begin{align} \label{eq:RG-model}
\cL =&\, i\sum_{j=1}^L\left[\Omega+\omega_j\right]\s^z_j  - \g_{0}\sum_{j,k=1}^L \s^z_j \s^z_k \nonumber\\
& \quad - \g \sum_{j,k=1}^L \sqrt{\omega_j\omega_k}\left(\s^x_j \s^x_k+\s^y_j \s^y_k\right),
\end{align}
where the $\s^{\alpha}$ are now spin-1 operators. Here we use $L$ to denote the total number of triplet states, and take $\omega_1 \dots \omega_L$ to denote the corresponding amplitudes in the Liouvillian.

This operator is clearly no longer Hermitian, but can be interpreted as a Richardson-Gaudin model with factorizable interaction and complex interaction constant, which however does not preclude an exact solution. The (right) eigenstates are a direct generalization of the eigenstates for the Hermitian model (see Appendix~\ref{app:RGModel}), and can be written as Bethe ansatz states of the form 
\begin{align}\label{eq:BetheState}
\ket{v_1 \dots v_N} = \prod_{a=1}^N\left(\sum_{j=1}^L \frac{\sqrt{\omega_j}}{\omega_j-v_a}\s_j^+\right)\ket{\emptyset},
\end{align}
where $\ket{\emptyset} = \ket{1,-1}_1 \otimes \dots \otimes \ket{1,-1}_L$ is a vacuum state annihilated by all $\s_j^-$, and the wave function is parametrized by a set of (possibly complex) parameters $\{v_1 \dots v_N \}$, also known as rapidities, satisfying the Bethe equations
\begin{align}\label{eq:BAE}
 - \frac{\g+i}{2\g} + \sum_{j=1}^L \frac{\omega_j}{\omega_j-v_a}-\sum_{b\neq a}^N\frac{ v_b}{v_b-v_a} = 0, \quad  a=1\dots N.
\end{align}
The Liouvillian has corresponding eigenvalues $\gamma(v_1,\dots,v_N)$, with
\begin{align}\label{eq:eigvalue_L}
\gamma(v_1,\dots,v_N) =&\,  (\g + i)\left(\sum_{a=1}^N v_a - \sum_{j=1}^L \omega_j \right) \nonumber\\
&\quad+i \Omega (N-L)- \g_0 (N-L)^2\,.
\end{align}

\section{Spectrum}
\label{sec:spectrum}
Before analyzing the eigenspectrum of the Liouvillian~\eqref{eq:RG-model}, it is useful to discuss its symmetries. First, it is clear that $\cL$ conserves total spin-$z$ projection, i.e.
\begin{align}
[\cL\,,\s^z] = 0, \qquad \s^z = \sum_j \s^z_j\,.
\end{align}
Undoing the mapping from operators to states, acting with $\s^z$ on a state corresponds to commuting the operator with $\sum_j s_j^z$.
All eigenstates of the Liouvillian are common eigenstates of $\s^z$, and it can easily be seen that
\begin{align}
\s^z \ket{v_1\dots v_N} = (N-L)\ket{v_1\dots v_N}\,.
\end{align}
This implies that we can set $\g_0 = \Omega = 0$ without loss of generality, since these simply correspond to a constant shift in the real and imaginary part of the eigenvalues respectively [see Eq.~\eqref{eq:eigvalue_L}]. This symmetry can also be observed from the Bethe equations \eqref{eq:BAE}, which are independent of both $\g_0$ and $\Omega$. 

Second, and crucially, the Liouvillian is pseudo-Hermitian. Namely, 
\begin{align}\label{eq:PTtransform}
\cL^\dagger = P^{\dagger}\cL P \quad \textrm{with} \quad P = \prod_{j=1}^L \s_j^x\,,
\end{align}
which follows from the observation that the non-Hermitian part $i\sum_j \omega_j \s_j^z$ maps to $-i \sum_j \omega_j \s_j^z$ under Hermitian conjugation, which can be undone through the spin inversion operator $P$, whereas the remaining parts of the Liouvillian are both Hermitian and invariant under spin inversion.

This can also be interpreted as PT symmetry, where a non-Hermitian Hamiltonian is invariant under a combined unitary (spin inversion) and antiunitary (complex conjugation) transformation, both of which square to identity \cite{bender_real_1998,prosen_p_2012}. On the level of the Liouvillian, spin inversion maps the Lindblad operator $L_+$ to $L_-$ and vice versa, exchanging the role of gain and loss. 
While the definition of PT symmetry for a Liouvillian is more involved than that for a non-Hermitian Hamiltonian, as discussed in e.g. Refs.~\cite{prosen_p_2012,huber_emergence_2020}, the mapping to a non-Hermitian model allows us circumvent these subtleties.

The above identity \eqref{eq:PTtransform} implies that left and right eigenstates are related through spin inversion. Furthermore, pseudo-Hermiticity implies that all eigenvalues either appear as part of a complex conjugate pair \cite{mostafazadeh_pseudo-hermiticity_2002,mostafazadeh_pseudo-hermitian_2010}, in which case the two eigenstates are related through the corresponding $PT$ transformation, or as purely real, in which case the eigenstate is invariant under $PT$. Since $P^{\dagger}\s^z P = - \s^z$, only eigenstates with $\s^z = 0$ (or $N=L$) can be invariant under $P$. In terms of operators, such states where $\s^z = 0$ correspond to the parts of the density matrix that preserve total spin $\sum_j s_j^z$ projection. This also implies that any eigenstate with nonzero $\s^z$ will have a complex eigenvalue, and its complex conjugate corresponds to an eigenstate with $-\s^z$.

\begin{figure}[tb!]                      
 \begin{center}
 \includegraphics[width=\columnwidth]{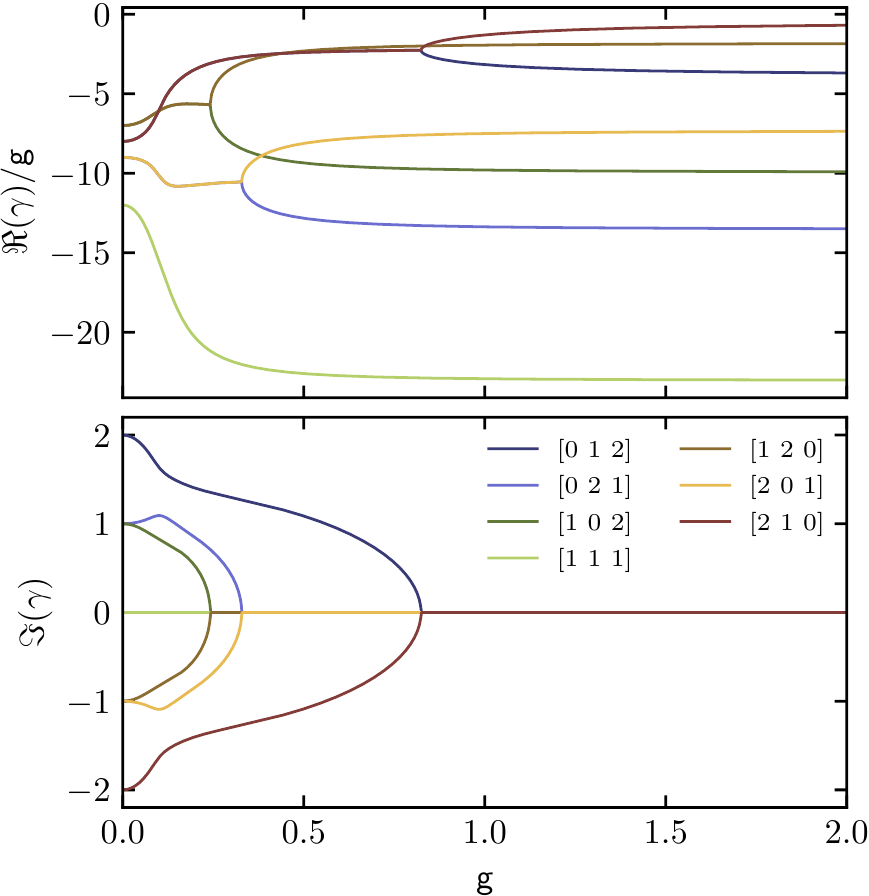}
 \caption{Eigenspectrum for $L=N=3$ with $\omega_i = i, i=1 \dots L$ as $\g$ is varied. Note that the real part is rescaled by $\g$. The eigenstates are labeled by the spin occupation numbers $[n_1\,n_2\,n_3]$ in the non-interacting limit $\g=0$, with $N = \sum_j n_j$. 
 \label{fig:smallsystem}}
 \end{center}
\vspace{-\baselineskip}
\end{figure}

In Fig.~\ref{fig:smallsystem}, we plot the eigenspectrum for a small system of $L=3$ spins as $\g$ is increased from zero to some nonzero final value, focusing on the symmetry sector $\s^z=0$. 
For concreteness, we consider a `picket-fence' model of evenly spaced levels $\omega_i = i, i=1\dots L$ \cite{hirsch_fully_2002}, although all results in the following are independent of the specific model unless explicitly mentioned. For $\g=0$ the Lindblad operators vanish and the spectrum is purely imaginary, as expected, and all eigenvalues are either zero or part of a complex conjugate pair $\pm(E_m-E_n)i$, with $E_{m,n}$ eigenvalues of the non-interacting Hamiltonian $H$ from Eq.~\eqref{eq:model}. As the coupling to the environment $\g$ is turned on, the eigenvalues acquire a real part. Further increasing $\g$, the imaginary parts of the complex conjugate eigenvalues coalesce and vanish -- the dynamics becoming purely dissipative once all eigenvalues have collapsed on the real line. At large $\g$, these purely real eigenvalues are then proportional to $\g$.

\begin{figure}[tb!]                      
 \begin{center}
 \includegraphics[width=\columnwidth]{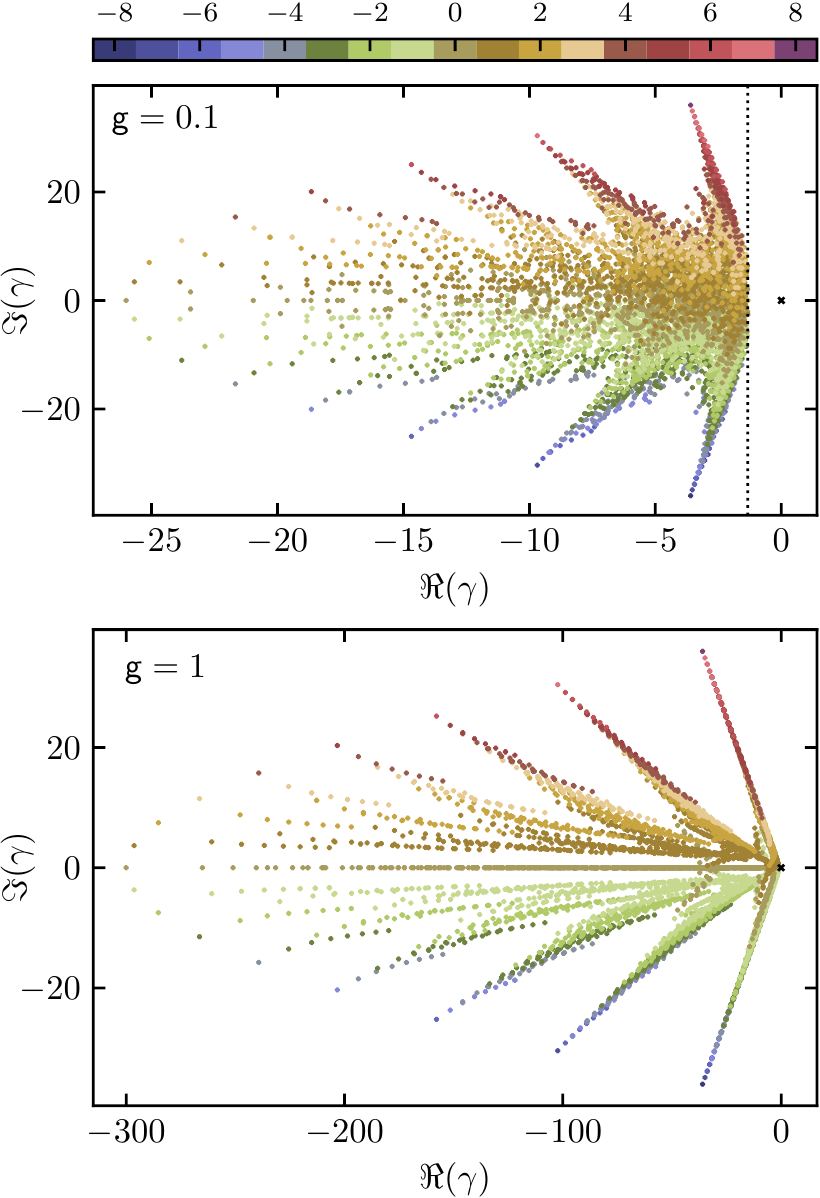}
 \caption{Eigenspectrum for $\omega_i = i, i=1 \dots L$ for fixed $\g$ and $L=8$. The eigenstates are color coded according to $\s^z$. For $\g=0.1$ the dotted vertical line indicates the spectral gap $\approx -1.328$ corresponding to a pair of complex conjugate states, for $\g=1$ the spectral gap $\approx -0.674$, and the cross indicates the origin as reference.
 \label{fig:scatter}}
 \end{center}
\vspace{-\baselineskip}
\end{figure}

This behavior readily extends to larger system sizes. In Fig.~\ref{fig:scatter} we plot the full spectrum for a system of $L=8$ spins at two different values of the coupling strength $\g$. The spectrum is symmetric with respect to the real axis because of the pseudo-Hermiticity, and symmetric states have opposite values of $\s^z$. Different limiting behaviors can be observed: at small $\g$ almost all states correspond to complex conjugate pairs, whereas for larger $\g$ the vast majority of states with $\s^z = 0$ have collapsed onto the real axis. If we would further increase $\g$ all states in this symmetry sector eventually become real. In both limits there is a nonvanishing spectral gap and a `continuum' of nearby states. However, for small $\g$ this gap is determined by a pair of complex conjugate eigenstates, whereas at large $\g$ the spectral gap is determined by a single real and nondegenerate eigenvalue. While the real part of most states is proportional to $\g$, the spectral gap in fact decreases after passing through an exceptional point (as can also be observed in Fig.~\ref{fig:smallsystem}).

In terms of the dynamics, a smaller spectral gap corresponds to a slower decay rate. For small $\g$ the leading decay mode will exhibit oscillations with a frequency set by the imaginary part of the leading eigenvalues, whereas for larger $\g$ the leading mode is purely dissipative but (possibly) decaying at a slower rate. Exactly at the exceptional point where the complex conjugate eigenvalues coalesce, the Liouvillian is no longer diagonalizable and a two-dimensional Jordan block will appear in its Jordan block decomposition, leading to critical dynamics $te^{-|\gamma| t}$ with $\gamma$ the leading eigenvalue \cite{ashida_non-hermitian_2020}. The exceptional point is accompanied by a non-analytic behavior of this leading eigenvalue, leading to a dissipative quantum phase transition. This transition can also be interpreted as the spontaneous breaking of PT-symmetry/pseudo-Hermiticity, since for large $\g$ the leading mode is invariant under PT symmetry, whereas at small $\g$ the two leading modes are no longer invariant under PT symmetry, but rather get mapped to each other.

This behavior is even more pronounced for nonzero $\g_0$. In this case all eigenvalues acquire an additional shift of the real value $-\g_0(\s^z)^2 = -\g (N-L)^2$, such that the decay rate of all sectors with nonzero $\s^z$ is increased. Only the relevant $\s^z=0$ sector, in which the transition occurs, is left invariant by both a nonzero $\g_0$ and nonzero $\Omega$. The latter induces an additional shift in the complex value of all sectors with nonzero $\s^z$, leading to global shift in all oscillation frequencies. As such, while the transition only occurs in the symmetry sector $\s^z=0$, this sector generally contains the leading eigenvalue and for sufficiently large $\g_0$ all dynamics are determined purely by this sector, with all other sectors rapidly decaying. The other symmetry sectors still exhibit nontrivial behavior as $\g$ is increased, since it can be seen that these eigenvalues organize themselves on (approximately) straight lines in the spectrum where $\Im(\omega) \propto \Re(\omega)$. These lines will be discussed in more detail in Sec.~\ref{subsec:strongcoupling}. In the following, we will consider different limits of this model where the exact solution allows for additional insight.

\subsection{Homogeneous limit}
One limit where the exact solution is particularly simple is the limit where all $\omega_j$ are equal, i.e. $\omega_j = \omega, \forall j$. In this case the model can be recast in terms of total spin operators $\s_{\rm tot}^{\alpha} = \sum_j \s_j^{\alpha}$ as
\begin{align}
\cL &= i(\Omega+\omega)\s^z_{\rm tot}  - \g_{0} \left(\s^z_{\rm tot}\right)^2- \g \omega \left(\s^x_{\rm tot} \s^x_{\rm tot}+\s^y_{\rm tot} \s^y_{\rm tot}\right) \nonumber\\
&=i(\Omega+\omega)\s^z_{\rm tot}  - \g_{0} \left(\s^z_{\rm tot}\right)^2- \g \omega \left[\Big(\vec{\s}_{\rm tot}\Big)^2 - \Big(\s_{\rm}^z\Big)^2\right]\,.
\end{align}
The eigenstates immediately follow as degenerate multiplets expressed in $\ket{\s_{\rm tot}, \s^z_{\rm tot}} = \ket{S,M}$, with total spin $S$ and total spin projection $M$. The corresponding eigenvalues $\gamma_{S,M}$ are given by
\begin{align}
\gamma_{S,M} = i \omega M - \g \omega \left[S(S+1)-M^2\right]\,,
\end{align}
again setting $\g_0 = \Omega= 0$ for convenience. In fact, the same model is obtained as for the homogeneous limit of the XXX Richardson-Gaudin model studied in Ref.~\cite{rowlands_noisy_2018}, only now with an interaction constant $\g \omega$ instead of $\g$. The degeneracy of these eigenvalues follows from the total number of ways in which $L$ spin-one particles can be coupled to total spin $S$, as given by the Riordan numbers \cite{andrews_threedimensional_1977,bernhart_catalan_1999,rowlands_noisy_2018}. 

\begin{figure}[tb!]                      
 \begin{center}
 \includegraphics[width=\columnwidth]{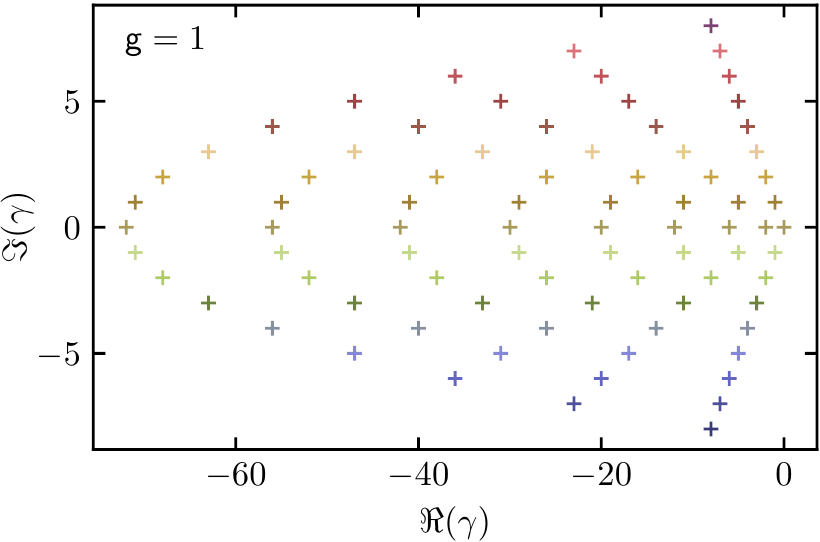}
 \caption{Eigenspectrum for the homogeneous model with $\omega_j = \omega = 1, \forall j$ for fixed $\g$ and $L=8$. The eigenstates are color coded according to $M$, as in Fig.~\ref{fig:scatter}.
 \label{fig:homogeneous}}
 \end{center}
\vspace{-\baselineskip}
\end{figure}

The spectrum is shown in Fig.~\ref{fig:homogeneous} for a system with $L = 8$ spins. The homogeneous model now has a nontrivial steady state within the triplet sector given by the state $\s_{\rm tot} = 0$, i.e. the fully isotropic state (see also Ref.~\cite{rowlands_noisy_2018}).
The real part is maximized when $M = \pm S$, and the leading eigenvalues are a pair of complex conjugate eigenvalues with $S = |M| = 1$, resulting in a spectral gap $\g \omega$. Furthermore, all states with $\s^z = M = 0$ are purely real at all values of $\g$, such that no transition occurs as $\g$ is varied.
In the $\s^z = 0$ sector of the homogeneous model, pseudo-Hermiticity is unbroken at all coupling strengths.
The nontrivial steady state and this lack of a transition is particular to the homogeneous model. 

More generally, note that the ratio of the real and imaginary part satisfies
\begin{eqnarray}
\g \frac{\Im(\gamma_{S,M})}{\Re(\gamma_{S,M})} = -\frac{M}{S(S+1)-M^2}\,,
\end{eqnarray}
independent of $\omega$. The real part is proportional to $\g$, whereas the imaginary part is independent of $\g$, reminiscent of Fig.~\ref{fig:scatter}.
Plugging in $M=\mp S$ returns a ratio $\pm 1$, plugging in $M = \mp(S-1)$ returns a ratio $\pm (S-1)/(3S-1) \approx \pm 1/3$ in the limit of large $S$, etc. This is already indicative of the lines observed in the bottom panel of Fig.~\ref{fig:scatter}, which will be shown to be smoothly connected to the solutions with fixed $S-M$.

\subsection{Strong-coupling limit}
\label{subsec:strongcoupling}

Part of the structure of the homogeneous model is recovered in the strong-coupling limit where $\g$ is sufficiently large. The clearest connection is that all eigenvalues for $\s^z=0$ collapse to the real line, resulting in an unbroken pseudo-Hermitian phase within this symmetry sector. 
In this limit we can also treat the non-Hermitian contribution to $\cL$ as a perturbation on top of the Hermitian interaction part. 
While there is no longer a nontrivial steady state, away from the homogeneous limit we can consider the behavior of the leading eigenvalue.
For $\g$ sufficiently small we observed that the leading eigenvalue belongs to a pair of complex conjugate states, whereas for $\g$ sufficiently large the leading eigenvalue is purely real and nondegenerate. 
As one particular application, in the latter limit the leading mode can be considered a perturbative correction of the ground state of the related Hermitian Hamiltonian
\begin{align}\label{eq:HermitianHam}
H = \sum_{j=1}^L\omega_j\s^z_j + \g \sum_{j,k=1}^L \sqrt{\omega_j\omega_k}\left(\s^x_j \s^x_k+\s^y_j \s^y_k\right)\,.
\end{align}
In the homogeneous limit, the ground state of this model is clearly given by the nontrivial steady state with $S_{\rm tot} = 0$, and we can now check what happens to this state in the inhomogeneous picket-fence model.

In order for the Bethe approach to be advantageous, we need to be able to explicitly target states of interest, e.g. the leading decay mode. 
However, the leading mode at (very) small coupling is not adiabatically connected to the leading mode at strong coupling, as also clear from Fig.~\ref{fig:smallsystem}. This is consistent with the Hermitian model, which undergoes a phase transition at $\g=1$. 
However, we observe that the ground state at large coupling in the Hermitian model is adiabatically connected to the leading mode of the non-Hermitian model at strong coupling. Empirically, we find that this ground state is connected to the state $(\s_1^{+})^2 (\s_3^{+})^2 (\s_5^{+})^2 \dots \ket{\emptyset}$ as $\g \to 0$ and to a nontrivial steady state with $\s_{\rm tot} = 0$ in the homogeneous limit, allowing us to explicitly find the spectral gap without having to iterate over all Bethe states.

\begin{figure}[tb!]                      
 \begin{center}
 \includegraphics[width=\columnwidth]{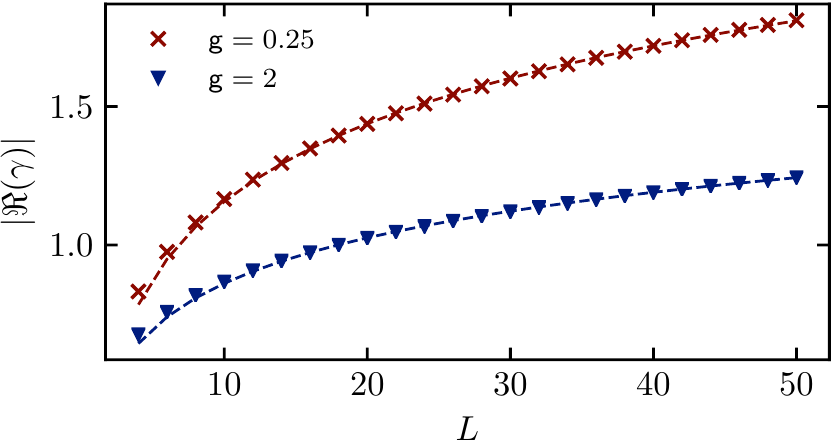}
 \caption{Absolute value of the real part of the leading eigenvalue for $\omega_i = i, i=1 \dots L$ with fixed $\g$ and varying $L$. Markers indicate exact results, dashed line is a logarithmic fit $\Re(\gamma) \propto \log L + \textrm{Cst}$. For $\g = 0.25$ the leading eigenvalue is part of a complex conjugate pair, whereas for $\g=2$ the leading eigenvalue is purely real and nondegenerate.
 \label{fig:scaling}}
 \end{center}
\vspace{-\baselineskip}
\end{figure}

Using this correspondence, we compare the scaling with system size of the real part of the leading eigenvalue for the picket-fence model at two values of $\g$ in Fig.~\ref{fig:scaling}. We have chosen $\g = 0.25$ to be small enough that the leading eigenvalue is still part of a complex conjugate pair at all considered system sizes, but large enough that we are away from the weak-coupling limit in which the eigenvalues can be treated perturbatively. Similarly, we have chosen $\g = 2$ to be large enough that the leading eigenvalue is real and proportional to $\g$, but small enough that it is clear that the spectral gap decreases after the exceptional point: for all considered system sizes, the spectral gap at $\g=2$ is smaller than the spectral gap at $\g=0.25$. The behavior of the (exact) spectral gap as system size $L$ is varied suggests a logarithmic scaling 
\begin{equation}
\Re(\gamma) \propto \log L + \textrm{Cst.}
\end{equation}
and hence only a slow growth as the system size is increased. Comparing the exact results with a logarithmic fit in Fig.~\ref{fig:scaling}, the correspondence is excellent for $L \geq 10$ until the maximal calculated value $L=50$. The Bethe ansatz allowed us to obtain exact results for a system of $L=50$ spins, where the full Hilbert space has dimension $3^{50} \approx 7.2 \cdot 10^{23}$.

Moving to the other symmetry sectors, we observe that the eigenvalues are approximately situated on straight lines where $\g \Im(\gamma)/\Re(\gamma) = \pm 1/(2n+1),n \in \mathbbm{Z}$. These can be seen as a remnant from the homogeneous limit, where we already noted that such a quantization occurs. From the Bethe ansatz, we can now generalize this behavior to general inhomogeneous models.

The rapidities solving the Bethe equations~\eqref{eq:BAE} can always be subdivided in two classes: the rapidities that are on the same order of magnitude as the $\omega_j$, and the rapidities that are large compared to all $\omega_j$. Assuming there are $p$ such large rapidities, we can denote the former as $v_a, a =1\dots q$, and the latter as $w_a, a=1 \dots p$, with $N = p+q$.
The states with $p$ large rapidities roughly correspond to the states with $M = -S+p$ in the homogeneous limit, since plugging in the large rapidities in the Bethe state \eqref{eq:BetheState} results in generalized raising operators of the form
\begin{equation}
\sum_{j=1}^L \frac{\sqrt{\omega_j}}{\omega_j-w_a}\s_j^+ \approx -\frac{1}{w_a} \sum_{j=1}^L\sqrt{\omega_j}\,\s_j^+,
\end{equation}
which reduces to a total spin raising operator in the homogeneous limit where $\omega_j = \omega, \forall j$.

The Bethe equations~\eqref{eq:BAE} for these large rapidities can similarly be approximated as
\begin{align}\label{eq:BAE_w}
 - \frac{\g+i}{2\g} - \frac{1}{w_a} \left( \sum_{k=1}^L \omega_k - \sum_{c=1}^{q} v_c \right) -\sum_{b\neq a}^p\frac{w_b}{w_b-w_a} \approx 0.
\end{align}
Multiplying this equation with $w_a$ and summing over $a=1\dots p$, the antisymmetric term drops out and we find (see also Appendix \ref{app:heinestieltjes}) that
\begin{align}
 \frac{\g+i}{2\g} \sum_{a=1}^p w_a \approx p \left(\sum_{c=1}^{q} v_c-\sum_{k=1}^L \omega_k  \right)\,.
\end{align}
Plugging this in the expression for the eigenvalue of $\cL$ [Eq.~\eqref{eq:eigvalue_L}], we find that the total eigenvalue $\gamma$ can be written as
\begin{align}
\gamma \approx -\left[(2p+1)\g + i\right]\left(\sum_{c=1}^{q} v_c - \sum_{k=1}^L \omega_k\right)\,.
\end{align}
In the Bethe equations for the remaning `finite' rapidities the dependence on $w_1, \dots, w_p$ simply drops out. In the strong-coupling limit these equations are approximately the Bethe equations for the Hermitian model, and finite rapidities can be treated as perturbative solutions to the Bethe equations for the Hermitian model (which is not possible for large rapidities). Furthermore, in the Hermitian model the rapidities only appear as real or as part of a complex conjugate pair, such that their sum is always real -- the Hermitian model necessarily has real eigenvalues. In this way we recover the previously observed quantization: for eigenstates of the Liouvillian \eqref{eq:RG-model} that can be approximately written as
\begin{align}
\left(\sum_{j=1}^L\sqrt{\omega_j}\,\s_j^+\right)^p \ket{v_1 \dots v_{q}},
\end{align}
the eigenvalues satisfy $\g {\Im(\gamma)}/{\Re(\gamma)} \approx 1/(2p+1)$. For these modes, the decay rate is approximately proportional to the oscillation frequency.

\section{Conserved charges}
\label{sec:charges}

In practice, Bethe equations of the form \eqref{eq:BAE} are rarely solved directly since they are plagued by singularities \cite{de_baerdemacker_richardson-gaudin_2012}. 
Rather, it is always possible to find a set of operator identities for the conserved charges of the integrable model, and these identities can be directly solved to obtain the eigenvalues of the conserved quantities and the integrable model.

While the interpretation of the conserved charges is partly lost in the non-Hermitian case, this formalism can be directly extended to the current case.
The Liouvillian still belongs to an extensive set of mutually commuting operators $[\cL,\Q_j] = 0, \forall j$ and $[\Q_j,\Q_k]=0, \forall j,k$, defined as
\begin{align}
\Q_j =\, & i\, \s_j^z + \g \left(\s_j^z\right)^2- 2 \g \sum_{k \neq j}^L \frac{\omega_k}{\omega_j-\omega_k}\s_j^z \s_k^z  \nonumber\\
&\quad - 2 \g \sum_{k \neq j}^L \frac{\sqrt{\omega_j \omega_k}}{\omega_j-\omega_k}\left(\s_j^x \s_k^x+\s_j^y \s_k^y\right)\,.
\end{align}
These are again a direct extension of the conserved quantities in the Hermitian model, as outlined in Appendix~\ref{app:RGModel}, and are clearly non-Hermitian.
Rather, these commuting quantities exhibit the same pseudo-Hermiticity/PT symmetry of the Liouvillian.
The Bethe ansatz states are common eigenstates of all $\Q_j$, and the corresponding eigenvalues $\q_j$ can be expressed in terms of the rapidities as
\begin{align}\label{eq:eigvalue_qi}
\q_j = (\g-i) - 2\g \sum_{a=1}^N \frac{v_a}{\omega_j-v_a}+ 2\g \sum_{k \neq j}^L \frac{\omega_k}{\omega_j-\omega_k} \,,
\end{align}
where we have made the dependence on the rapidities implicit. 
Note that $\cL$ is not linearly independent of these operators, since 
\begin{align}
\sum_{j=1}^L \omega_j \Q_j =&\, i \sum_j \omega_j \s_j^z + \g \sum_j \omega_j \left(\s_j^z\right)^2 \nonumber\\
&- \g \sum_{j} \sum_{k \neq j} \sqrt{\omega_j \omega_k} \left(\s_j^x \s_k^x+\s_j^y \s_k^y\right) \nonumber\\
=&\, \cL + \g \sum_{j} \omega_j \left[\left(\s_j^x\right)^2+\left(\s_j^y\right)^2+\left(\s_j^z\right)^2\right]\,.
\end{align}
The final expression has an additional contribution from the Casimir operators of the spin-1 operators, which can be treated as a constant.
The same relation holds for the eigenvalues of $\cL$ and $\Q_j$, since all operators can be simultaneously diagonalized, and we find that the eigenvalues $\gamma$ of $\cL$ can be expanded as
\begin{align}
\gamma = \sum_{j=1}^L \omega_j\, (\q_j - 2\g) \,.
\end{align}
We can similarly recover the conservation of total $\s^z = \sum_j \s_j^z$ from
\begin{equation}
\sum_{j=1}^L \Q_j = i\, \s^z + \g \left(\s^z\right)^2\,.
\end{equation}
\begin{figure}[tb!]                      
 \begin{center}
 \includegraphics[width=\columnwidth]{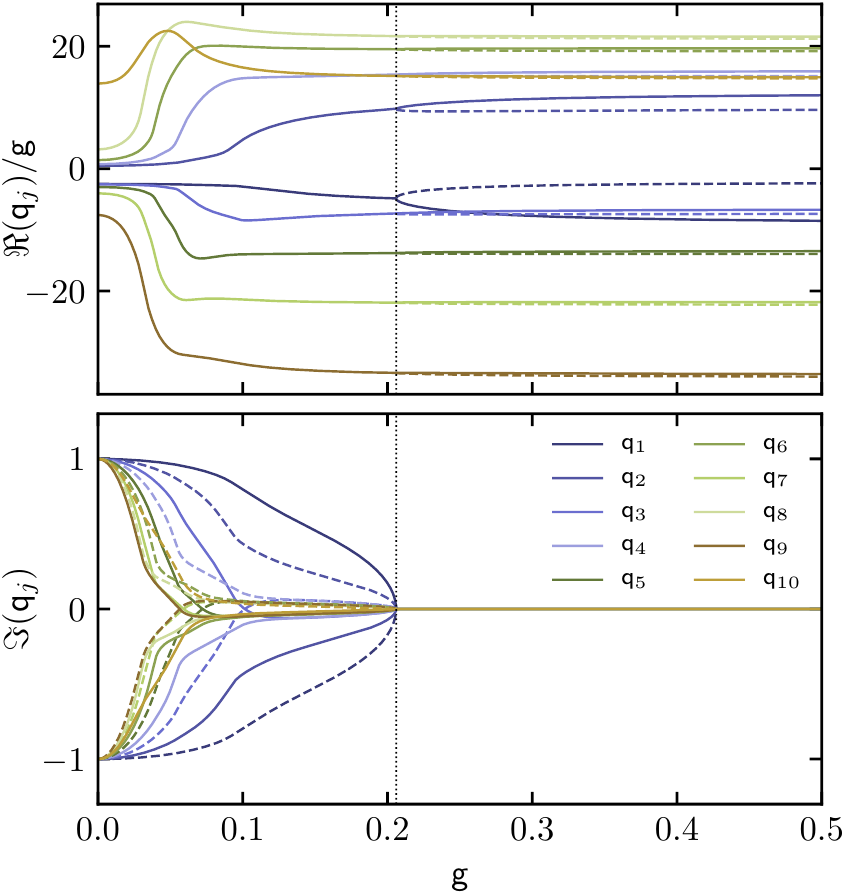}
 \caption{Real and imaginary part of the eigenvalues $\q_j$ of the commuting operators $\Q_j$ with $\omega_j = j, j=1 \dots L$ for $L=10$ and varying $\g$. Full lines indicate the state that is adiabatically connected to the leading mode in the strong-coupling limit, dashed lines indicate the state that is related to this state by PT transformation. The transition is marked by a vertical dotted line, and before the transition the eigenvalues are related by complex conjugation, while after the transition the eigenvalues are purely real and no longer related. Note that the real part is rescaled by $\g$.
 \label{fig:eigenvalues}}
 \end{center}
\vspace{-\baselineskip}
\end{figure}

Any set of rapidities solving the Bethe equations \eqref{eq:BAE} determines a single Bethe state \eqref{eq:BetheState}, and this state will be a common eigenstate of all conserved charges with eigenvalues $\{\q_1, \q_2 \dots \q_L\}$. Rather than first solving the Bethe equations for the rapidities, it is now possible to find a set of equations directly returning this set of eigenvalues, avoiding the explicit use of rapidities.
This approach has the advantage that the equations that need to be solved do not display the singular behavior of the regular Bethe equations \eqref{eq:BAE}. These equations can be derived using the approach from Ref.~\cite{claeys_eigenvalue-based_2015} and are explicitly derived in Appendix~\ref{app:EVB}. Defining shifted eigenvalues
\begin{align}
\gamma_j = \q_j + \g + 2\g \sum_{k \neq j}^L\frac{\omega_j}{\omega_j-\omega_k},
\end{align}
these satisfy the set of equations
\begin{align}\label{eq:EVB}
&\left[\gamma_j - 2\g \sum_{k \neq j}^L \frac{\omega_j}{\omega_j - \omega_k}\right]\left[\gamma_j^2 + \beta^2 - 4  \g \omega_j \sum_{k \neq j}^L\frac{\gamma_j - \gamma_k}{\omega_j - \omega_k}\right] \nonumber\\
&\qquad \quad = \g (\gamma_j^2 + \beta^2) - 4 \g^2 \omega_j^2 \sum_{k \neq j}^L\frac{\gamma_j - \gamma_k}{(\omega_j - \omega_k)^2},
\end{align}
with $\beta = 1+ 2 i \g (L-N)$. Taking the complex conjugate of these equations maps $\beta \to \beta^*$, and $\beta$ is purely real if $\s^z=0$ and hence $N=L$, reflecting the pseudo-Hermiticity in this sector. For $\s^z \neq 0$ the imaginary part gets mapped to $L-N \to N-L = L - (2 L - N)$, connecting the sectors with $N$ and $2L-N$ spin excitations and hence opposite values of $\s^z$.

Since these hold for all eigenvalues and all considered operators can be simultaneously diagonalized, the operators themselves satisfy the same set of coupled cubic equations.
These equations in fact form the backbone of our numerical approach.
For $\g=0$ the equations decouple and we find $\gamma_j (\gamma_j^2 + 1) = 0, \forall j$, which can be solved as $\gamma_j = 0$ or $\gamma_j = \pm i$, returning the expected eigenvalues of $i \s_j^z$. Solutions at nonzero $\g$ can be obtained by slowly increasing $\g$ to its final value and iteratively solving the equations at intermediate values of $\g$ using the solutions at smaller $\g$ as starting point.
In this way the solutions of Eq.~\eqref{eq:EVB} can be directly connected to the occupation numbers in the non-interacting limit, providing a way of targeting specific states (see e.g. Ref.~\cite{claeys_eigenvalue-based_2015} for details).
Such an approach is common in Richardson-Gaudin models \cite{babelon_bethe_2007,faribault_gaudin_2011,el_araby_bethe_2012,claeys_eigenvalue-based_2015,claeys_read-green_2016,claeys_inner_2017,claeys_richardson-gaudin_2018,dimo_quadratic_2018,faribault_bethe-ansatz-free_2018,claeys_integrable_2019}. Once these eigenvalues are known the rapidities can either be extracted from Eq.~\eqref{eq:eigvalue_qi} or the states themselves can be directly expressed in terms of these eigenvalues \cite{claeys_inner_2017,faribault_bethe-ansatz-free_2018}.
This equivalence can also be seen as a version of the generalized eigenstate thermalization hypothesis \cite{dalessio_quantum_2016}, since all Bethe states are fully determined by the associated conservation laws. 

In terms of the non-Hermitian model considered so far, this has two important consequences. First, we note that the pseudo-Hermiticity-breaking transition can be observed not just in the eigenspectrum of $\cL$, but also in the spectrum of all $\Q_j$. 
This is illustrated in Fig.~\ref{fig:eigenvalues} for a pair of representative eigenstates.
At the transition, the corresponding eigenvalues of all commuting operators change from complex conjugate to purely real. Second, since the states are completely determined by this set of eigenvalues and these eigenvalues are identical at the transition, the states themselves are identical and coalesce: the transition is accompanied by an exceptional point and not an accidental degeneracy.

\section{Conclusion} 
\label{sec:conc}
We discussed the exact solution of a Liouvillian with collective dissipation through a mapping to a non-Hermitian Richardson-Gaudin model. The resulting Hamiltonian is pseudo-Hermitian/PT-symmetric, and as the coupling to the environment is increased the eigenvalues change from complex conjugate pairs to purely real. Such a transition is accompanied by a dissipative phase transition and an exceptional point in the spectrum of the Liouvillian, reminiscent of the quantum phase transition in the corresponding Hermitian model. In this way we find an exactly solvable model for an open and interacting quantum system exhibiting nontrivial dynamics.

In the limit where the model is fully homogeneous the eigenspectrum can be expressed in terms of total spin quantum numbers and supports a nontrivial steady state. Away from this limit the decay rates can be analyzed using the exact Bethe ansatz solution, where we find that the nontrivial steady state now decays with a decay rate that increases slowly (logarithmically) with system size. The Bethe ansatz approach is crucial in establishing the logarithmic scaling, since exact eigenvalues can be obtained for system sizes where the Hilbert space is too large for traditional exact methods.
For higher excited states in different symmetry sectors we observe that the decay rate is proportional to the oscillation frequency with a quantized prefactor, which is reflected in the eigenvalues organizing themselves in approximately straight lines in the complex plane.

\section*{Acknowledgements} 
We gratefully acknowledge support from EPSRC Grant No. EP/P034616/1.  We thank Jan Behrends for useful discussions.

\appendix
\section{Hermitian model}
\label{app:RGModel}
In this Appendix we provide an overview of the exact solution of the Hermitian XXZ Richardson-Gaudin model (see e.g. \cite{ortiz_exactly-solvable_2005,ibanez_exactly_2009,rombouts_quantum_2010,van_raemdonck_exact_2014,links_exact_2015,claeys_read-green_2016,claeys_richardson-gaudin_2018}). For a set of $L$ spin-1 particles, the conserved quantities $Q_j, j=1\dots L$ are given by
\begin{align}
Q_j = S_j^z + g \sum_{k \neq j}^L\left[\frac{\sqrt{\omega_j \omega_k}}{\omega_j-\omega_k}\left(S_j^+ S_k^-+S_j^- S_k^+\right)+ \frac{\omega_j + \omega_k}{\omega_j-\omega_k}S_j^z S_k^z \right],
\end{align}
satisfying $[Q_j,Q_k] = 0, \forall j,k$. The Bethe states are defined as
\begin{equation}
\ket{v_1 \dots v_N} = \prod_{a=1}^N\left(\sum_{j=1}^L \frac{\sqrt{\omega_j}}{\omega_j-v_a}S_j^+\right)\ket{\emptyset},
\end{equation}
with $\ket{\emptyset} = \ket{1,-1}_1 \otimes \dots \otimes \ket{1,-1}_L$, and the Bethe equations are given by
\begin{equation}
\frac{1}{g} + \sum_{j=1}^L \frac{\omega_j+v_a}{\omega_j-v_a} -  \sum_{b \neq a}^N \frac{v_b+v_a}{v_b-v_a} = 0, \quad a =1 \dots N,
\end{equation}
with corresponding eigenvalues
\begin{equation}
q_j = - \left[1+g \sum_{a=1}^N \frac{\omega_j+v_a}{\omega_j-v_a}-g \sum_{k \neq j}^L \frac{\omega_j+\omega_k}{\omega_j-\omega_k} \right]\,.
\end{equation}

A Hamiltonian with all-to-all interactions can be constructed by writing
\begin{align}
\sum_j \omega_j (Q_j + g \cC_j) =& \sum_j \omega_j S_j^z \left(1+g \sum_k S_k^z \right) \nonumber\\
&+ g \sum_{j,k}\sqrt{\omega_j \omega_k}\left(S_j^x S_k^x + S_j^y S_k^y\right)\,,
\end{align}
where $\cC_j = (S_j^x)^2+ (S_j^y)^2+ (S_j^z)^2 = 2$ is the Casimir operator for each algebra. Note that part of the interaction has been absorbed in the first term, but since total spin-$z$ projection is a conserved quantity it can be replaced by its eigenvalue. We can define a new interaction strength
\begin{align}
G^{-1} = g^{-1} + \sum_j S_j^z = g^{-1}+(N-L)\,,
\end{align}
and we obtain the Hermitian Hamiltonian from the main text as
\begin{align}
&\sum_i \omega_j (Q_j + g \cC_j)/(1+g \sum_k S_k^z) \nonumber\\ 
&\qquad=\sum_j \omega_j S_j^z + G \sum_{j,k}\sqrt{\omega_j \omega_k}\left(S_j^x S_k^x + S_j^y S_k^y\right)\,.
\end{align}
The Bethe equations can be rewritten in terms of this new coupling constant as
\begin{align}
\frac{1-G}{2G}+ \sum_{j=1}^L \frac{\omega_j}{\omega_j-v_a}-\sum_{b\neq a}^N\frac{v_b}{v_b-v_a} = 0,
\end{align}
as well as the conserved charges
\begin{align}
&Q_j/(1+g \sum_k S_k^z) = S_j^z - G \left(S_j^z\right)^2 \nonumber\\
& + 2G\sum_{k \neq j}^L \left[\frac{\sqrt{\omega_j \omega_k}}{\omega_j-\omega_k}\left(S_j^x S_k^x+S_j^y S_k^y\right)+ \frac{\omega_k}{\omega_j-\omega_k}S_j^z S_k^z \right],
\end{align}
where the corresponding eigenvalues of this rescaled operator can be written as
\begin{align}
-(1+G) + 2G \sum_{a=1}^N \frac{v_a}{\omega_j-v_a}- 2G \sum_{k \neq j}^L \frac{\omega_k}{\omega_j-\omega_k}\,.
\end{align}
The Liouvillian from the main text now corresponds to a non-Hermitian XXZ model with $G= i \g$.

For completeness, we note that the Hermitian model undergoes a phase transition at $|G|=1$. This can easily be understood since we can write
\begin{align}
H &= \sum_j \omega_j S_j^z+\frac{G}{2}\left(Q^{\dagger}Q+QQ^{\dagger}\right) \nonumber\\
&= \frac{1}{2}[Q^{\dagger},Q]+\frac{G}{2}\left(Q^{\dagger}Q+QQ^{\dagger}\right)\,,
\end{align}
with $Q^{\dagger} = \sum_{j} \sqrt{\omega_j}\, S_j^{+}$. We can hence rewrite $H$ as
\begin{align}
H = \frac{G+1}{2} Q^{\dagger}Q + \frac{G-1}{2}Q Q^{\dagger}\,. 
\end{align}
For $|G|<1$ the ground state is adiabatically connected to the non-interacting ground state at $G=0$ i.e. $(S_1^{+})^2 (S_2^{+})^2 \dots (S_{N/2}^{+})^2 \ket{\emptyset}$ for $\omega_i < \omega_j$ if $i<j$ and $N$ even, filling the vacuum state with $N$ spin excitations. At $|G|=1$ there is a quantum phase transition for either $N \leq L$ (at $G=1$) or $N \geq L$ (at $G=-1$). At the transition the Hamiltonian is positive semi-definite, and the ground state has zero energy and is highly degenerate. E.g. for $G=1$ the Hamiltonian can be written as $Q^{\dagger}Q$, for which the ground states are the states annihilated by $Q$, also known as dark states \cite{villazon_integrability_2020}, and there is a combinatorial number of such states for $N \leq L$. For $G>1$ we find that the ground state is adiabatically connected to $(S_1^+)^2(S_3^+)^2(S_5^+)^2 \dots\ket{\emptyset}$. This transition can be interpreted as a transition to a collective phase, where the ground state reduces to a fully isotropic singlet state $S_{\rm tot} = 0$ in the homogeneous limit.

\section{Eigenspectrum for $\s^z \neq 0$}
\label{app:nonzeroSz}
\begin{figure}[tb!]                      
 \begin{center}
 \includegraphics[width=\columnwidth]{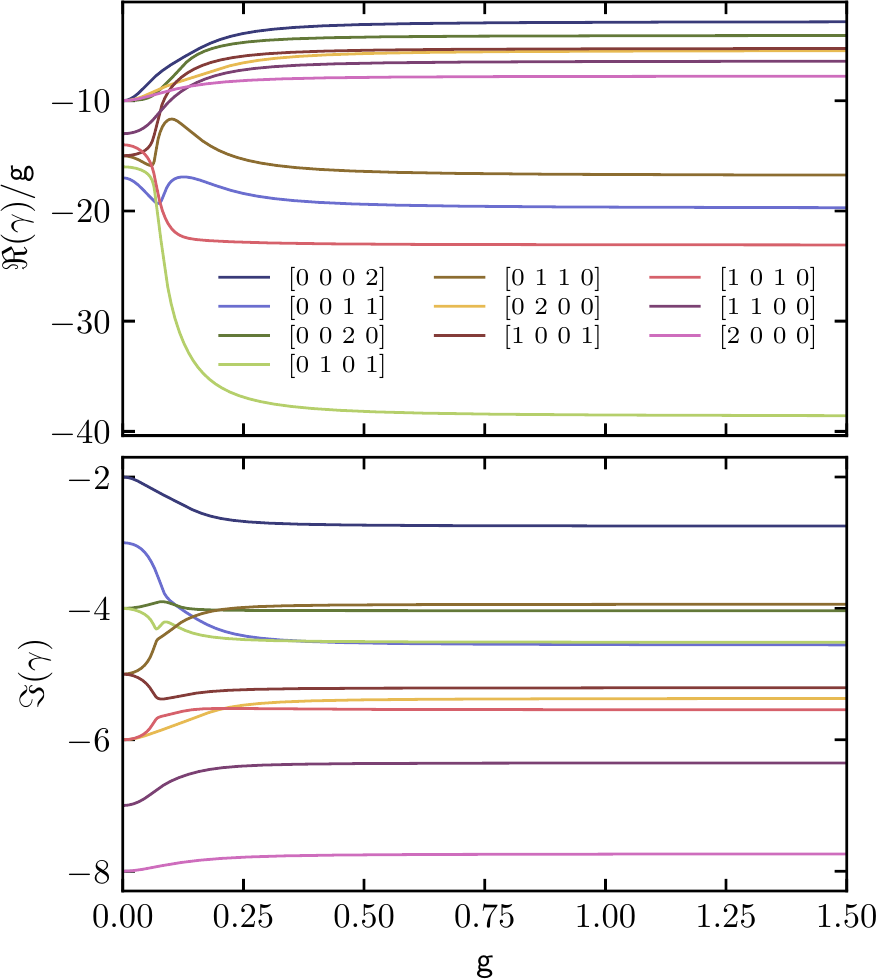}
 \caption{Eigenspectrum for $L=4$ and $N=2$ with $\omega_i = i, i=1 \dots L$ as $\g$ is varied. Note that the real part is rescaled by $\g$. The eigenstates are labeled by the spin occupation numbers $[n_1\,n_2\,n_3\,n_4]$ in the non-interacting limit $\g=0$, with $N = \sum_j n_j$. 
 \label{fig:smallsystem_nonzeroSz}}
 \end{center}
\vspace{-\baselineskip}
\end{figure}

For completeness, we show the eigenspectrum of the non-Hermitian Hamiltonian with $\s^z \neq 0$ in Fig.~\ref{fig:smallsystem_nonzeroSz}. All eigenvalues remain complex as $\g$ is increased, and at large coupling strengths the imaginary part is approximately constant whereas the real part is proportional to $\g$.

\section{Eigenvalues of the commuting operators}
\label{app:EVB}
In order to obtain the equations determining the eigenvalues of the commuting operators, we define a continuous function
\begin{align}
\Lambda(u) = \sum_{a=1}^N\frac{1}{u-v_a}\,
\end{align}
following Refs.~\cite{el_araby_bethe_2012,claeys_eigenvalue-based_2015}. We again start from the Hermitian model, for which the Bethe equations can be rewritten as
\begin{align}
\frac{\alpha}{2 v_a} + \sum_{j=1}^L \frac{1}{\omega_j-v_a}-\sum_{b\neq a}^N\frac{1}{v_b-v_a} = 0,
\end{align}
where we have introduced $\alpha = 1+ G^{-1}+2(L-N)$ for convenience. We then find
\begin{align}
&\left[\Lambda(u)\right]^2 = \sum_{a}\frac{1}{(u-v_a)^2}+\sum_a \sum_{b \neq a}\frac{1}{(u-v_a)}\frac{1}{(u-v_b)} \nonumber\\
&\quad=\sum_{a}\frac{1}{(u-v_a)^2}+2 \sum_a \sum_{b \neq a}\frac{1}{(u-v_a)}\frac{1}{(v_a-v_b)}  \nonumber\\
&\quad=\sum_{a}\frac{1}{(u-v_a)^2}-2 \sum_a \frac{1}{(u-v_a)}  \left[\frac{\alpha}{2 v_a} + \sum_{j} \frac{1}{\omega_j-v_a}\right] \nonumber\\
&\quad=\sum_{a}\frac{1}{(u-v_a)^2} - \frac{\alpha}{u}\sum_a  \left[\frac{1}{u-v_a}+\frac{1}{v_a}\right] \nonumber\\
&\qquad\quad-2 \sum_{j \neq i} \sum_a \frac{1}{\omega_j-u}\left[\frac{1}{u-v_a}-\frac{1}{\omega_j-v_a}\right] \nonumber\\
&\qquad\quad- 2 \sum_a \frac{1}{(u-v_a)(\omega_i-v_a)},
\end{align}
where we have performed a partial fraction decomposition in the second and fourth line, assuming $u \neq \omega_j, j \neq i$ for a fixed $i$, and used the Bethe equations to evaluate the summation $b\neq a$. These can be further evaluated since
\begin{align}
\sum_a \frac{\alpha}{2 v_a}  = - \sum_{j,a} \frac{1}{\omega_j-v_a}  = - \sum_j \Lambda(\omega_j),
\end{align}
again making use of the Bethe equations, to find
\begin{align}
\left[\Lambda(u)\right]^2 &=  -\frac{\alpha}{u}\Lambda(u) +\frac{2}{u}  \sum_j \Lambda(\omega_j)+2 \sum_{j \neq i}\frac{\Lambda(u)-\Lambda(\omega_j)}{u-\omega_j} \nonumber\\
& + \sum_{a}\frac{1}{(u-v_a)^2}- 2 \sum_a \frac{1}{(u-v_a)(\omega_i-v_a)}\,.
\end{align}
We can plug in $u=\omega_i$ and multiply the equation with $\omega_i$ to find
\begin{align}
\omega_i \Lambda(\omega_i)^2 =&  -\alpha\Lambda(\omega_i) +2\sum_j \Lambda(\omega_j)  -  \sum_a \frac{\omega_i}{(\omega_i-v_a)^2} \nonumber\\
&\qquad+2 \omega_i \sum_{j \neq i}\frac{\Lambda(\omega_i)-\Lambda(\omega_j)}{\omega_i-\omega_j}\,.
\end{align}
This is almost a closed set of equations for $\{\Lambda(\omega_1) \dots \Lambda(\omega_L)\}$, except for the dependence on $\sum_a 1/{(\omega_i-v_a)^2} = -\Lambda'(\omega_i)$. This dependence can be removed by taking the derivative of the above equations w.r.t. $u$ and again evaluating at $u = \omega_i$, leading to 
\begin{align}
&2 \Lambda(\omega_i) \Lambda'(\omega_i) = -\frac{\alpha}{\omega_i}\Lambda'(\omega_i)+ \frac{\alpha-2}{\omega_i^2}\Lambda(\omega_i) - \frac{2}{\omega_i^2} \sum_{j \neq i}\Lambda(\omega_j) \nonumber\\
&\qquad+ 2 \sum_{j \neq i}\frac{\Lambda'(\omega_i)}{\omega_i-\omega_j}-2\sum_{j \neq i}\frac{\Lambda(\omega_i)-\Lambda(\omega_j)}{(\omega_i-\omega_j)^2}\,.
\end{align}
This equation can be used to express $\Lambda'(\omega_i)$ in terms of $\{\Lambda(\omega_1) \dots \Lambda(\omega_L)\}$, which can then be plugged into the previously obtained equations to obtain a closed set of equations. Defining
\begin{align}
\gamma_i = 2 \omega_i G \Lambda(\omega_i) +2G(L-N)+ 1
\end{align}
then returns Eq.~\eqref{eq:EVB} after some straightforward manipulations.

\section{Heine-Stieltjes connection}
\label{app:heinestieltjes}
In this Appendix, we find an explicit expression for the large rapidities $w_1\dots w_p$ in terms of the `finite' rapidities $v_1 \dots v_q$, given Eq.~\eqref{eq:BAE_w}. Defining $\Delta = \sum_{j=1}^L \omega_j - \sum_{c=1}^q v_c$, we can write the Bethe equations for the large rapidities as
\begin{align}
 - \frac{\g+i}{2\g} - \frac{\Delta}{w_a} -\sum_{b\neq a}^p\frac{w_b}{w_b-w_a} = 0,\quad a=1\dots p.
\end{align}
We can relate the solutions of these equations to the roots of associated Laguerre functions. The associated Laguerre polynomials $L^{\alpha}_n(z)$ satisfy the differential equation
\begin{equation}
z P''(z)+(1+\alpha-z)P'(z)+n P(z) = 0,
\end{equation} 
for $P(z) = L^{\alpha}_n(z)$, and from the Heine-Stieltjes connection \cite{stieltjes_theoreme_1885,sriram_shastry_solution_2001} the roots $z_a, a=1\dots n$ are coupled through
\begin{equation}
1-\frac{1+\alpha}{z_a} - 2\sum_{b \neq a}^n \frac{1}{z_a-z_b} = 0,  \quad a=1\dots n.
\end{equation}
The Bethe equations for the large roots can be recast in terms of $x_a = 1/w_a$ as
\begin{align}
 - \frac{\g+i}{2\g}\frac{1}{x_a} - \Delta -\sum_{b\neq a}^p\frac{1}{x_a-x_b} = 0\,.
\end{align}
These are exactly the equations for the roots of the associated Laguerre polynomials $L^{\alpha}_n(z)$ with $z_a = -2 \Delta x_a = -2 \Delta/w_a$, $n=p$ and $\alpha = i/\g$. We can also recover the sum rule from the main text since
\begin{equation}
\sum_{a=1}^n \frac{1}{z_a} = -\frac{P'(0)}{P(0)}  = \frac{n}{1+\alpha}\,,
\end{equation}
which here reduces to
\begin{equation}
\sum_{a=1}^q w_a = -2 \Delta \frac{p}{1+i/\g}\,.
\end{equation}

\bibliography{Library_Open.bib}

\begin{thebibliography}{57}%
\makeatletter
\providecommand \@ifxundefined [1]{%
 \@ifx{#1\undefined}
}%
\providecommand \@ifnum [1]{%
 \ifnum #1\expandafter \@firstoftwo
 \else \expandafter \@secondoftwo
 \fi
}%
\providecommand \@ifx [1]{%
 \ifx #1\expandafter \@firstoftwo
 \else \expandafter \@secondoftwo
 \fi
}%
\providecommand \natexlab [1]{#1}%
\providecommand \enquote  [1]{``#1''}%
\providecommand \bibnamefont  [1]{#1}%
\providecommand \bibfnamefont [1]{#1}%
\providecommand \citenamefont [1]{#1}%
\providecommand \href@noop [0]{\@secondoftwo}%
\providecommand \href [0]{\begingroup \@sanitize@url \@href}%
\providecommand \@href[1]{\@@startlink{#1}\@@href}%
\providecommand \@@href[1]{\endgroup#1\@@endlink}%
\providecommand \@sanitize@url [0]{\catcode `\\12\catcode `\$12\catcode
  `\&12\catcode `\#12\catcode `\^12\catcode `\_12\catcode `\%12\relax}%
\providecommand \@@startlink[1]{}%
\providecommand \@@endlink[0]{}%
\providecommand \url  [0]{\begingroup\@sanitize@url \@url }%
\providecommand \@url [1]{\endgroup\@href {#1}{\urlprefix }}%
\providecommand \urlprefix  [0]{URL }%
\providecommand \Eprint [0]{\href }%
\providecommand \doibase [0]{https://doi.org/}%
\providecommand \selectlanguage [0]{\@gobble}%
\providecommand \bibinfo  [0]{\@secondoftwo}%
\providecommand \bibfield  [0]{\@secondoftwo}%
\providecommand \translation [1]{[#1]}%
\providecommand \BibitemOpen [0]{}%
\providecommand \bibitemStop [0]{}%
\providecommand \bibitemNoStop [0]{.\EOS\space}%
\providecommand \EOS [0]{\spacefactor3000\relax}%
\providecommand \BibitemShut  [1]{\csname bibitem#1\endcsname}%
\let\auto@bib@innerbib\@empty
\bibitem [{\citenamefont {Breuer}\ and\ \citenamefont
  {Petruccione}(2007)}]{breuer_theory_2007}%
  \BibitemOpen
  \bibfield  {author} {\bibinfo {author} {\bibfnamefont {H.-P.}\ \bibnamefont
  {Breuer}}\ and\ \bibinfo {author} {\bibfnamefont {F.}~\bibnamefont
  {Petruccione}},\ }\href
  {https://oxford.universitypressscholarship.com/view/10.1093/acprof:oso/9780199213900.001.0001/acprof-9780199213900}
  {\emph {\bibinfo {title} {The {Theory} of {Open} {Quantum} {Systems}}}}\
  (\bibinfo  {publisher} {Oxford University Press},\ \bibinfo {year}
  {2007})\BibitemShut {NoStop}%
\bibitem [{\citenamefont
  {Mostafazadeh}(2010)}]{mostafazadeh_pseudo-hermitian_2010}%
  \BibitemOpen
  \bibfield  {author} {\bibinfo {author} {\bibfnamefont {A.}~\bibnamefont
  {Mostafazadeh}},\ }\bibfield  {title} {\bibinfo {title} {Pseudo-{Hermitian}
  {Representation} of {Quantum} {Mechanics}},\ }\href
  {https://doi.org/10.1142/S0219887810004816} {\bibfield  {journal} {\bibinfo
  {journal} {Int. J. Geom. Methods Mod. Phys.}\ }\textbf {\bibinfo {volume}
  {07}},\ \bibinfo {pages} {1191} (\bibinfo {year} {2010})}\BibitemShut
  {NoStop}%
\bibitem [{\citenamefont {El-Ganainy}\ \emph {et~al.}(2018)\citenamefont
  {El-Ganainy}, \citenamefont {Makris}, \citenamefont {Khajavikhan},
  \citenamefont {Musslimani}, \citenamefont {Rotter},\ and\ \citenamefont
  {Christodoulides}}]{el-ganainy_non-hermitian_2018}%
  \BibitemOpen
  \bibfield  {author} {\bibinfo {author} {\bibfnamefont {R.}~\bibnamefont
  {El-Ganainy}}, \bibinfo {author} {\bibfnamefont {K.~G.}\ \bibnamefont
  {Makris}}, \bibinfo {author} {\bibfnamefont {M.}~\bibnamefont {Khajavikhan}},
  \bibinfo {author} {\bibfnamefont {Z.~H.}\ \bibnamefont {Musslimani}},
  \bibinfo {author} {\bibfnamefont {S.}~\bibnamefont {Rotter}},\ and\ \bibinfo
  {author} {\bibfnamefont {D.~N.}\ \bibnamefont {Christodoulides}},\ }\bibfield
   {title} {\bibinfo {title} {Non-{Hermitian} physics and {PT} symmetry},\
  }\href {https://doi.org/10.1038/nphys4323} {\bibfield  {journal} {\bibinfo
  {journal} {Nature Phys}\ }\textbf {\bibinfo {volume} {14}},\ \bibinfo {pages}
  {11} (\bibinfo {year} {2018})}\BibitemShut {NoStop}%
\bibitem [{\citenamefont {Huber}\ \emph {et~al.}(2020)\citenamefont {Huber},
  \citenamefont {Kirton}, \citenamefont {Rotter},\ and\ \citenamefont
  {Rabl}}]{huber_emergence_2020}%
  \BibitemOpen
  \bibfield  {author} {\bibinfo {author} {\bibfnamefont {J.}~\bibnamefont
  {Huber}}, \bibinfo {author} {\bibfnamefont {P.}~\bibnamefont {Kirton}},
  \bibinfo {author} {\bibfnamefont {S.}~\bibnamefont {Rotter}},\ and\ \bibinfo
  {author} {\bibfnamefont {P.}~\bibnamefont {Rabl}},\ }\bibfield  {title}
  {\bibinfo {title} {Emergence of {PT}-symmetry breaking in open quantum
  systems},\ }\href {https://doi.org/10.21468/SciPostPhys.9.4.052} {\bibfield
  {journal} {\bibinfo  {journal} {SciPost Phys.}\ }\textbf {\bibinfo {volume}
  {9}},\ \bibinfo {pages} {052} (\bibinfo {year} {2020})}\BibitemShut {NoStop}%
\bibitem [{\citenamefont {Nakanishi}\ and\ \citenamefont
  {Sasamoto}(2021)}]{nakanishi_pt_2021}%
  \BibitemOpen
  \bibfield  {author} {\bibinfo {author} {\bibfnamefont {Y.}~\bibnamefont
  {Nakanishi}}\ and\ \bibinfo {author} {\bibfnamefont {T.}~\bibnamefont
  {Sasamoto}},\ }\bibfield  {title} {\bibinfo {title} {{PT} phase transition in
  open quantum systems with {Lindblad} dynamics},\ }\href
  {http://arxiv.org/abs/2104.07349} {\bibfield  {journal} {\bibinfo  {journal}
  {arXiv:2104.07349 [cond-mat, physics:quant-ph]}\ } (\bibinfo {year}
  {2021})}\BibitemShut {NoStop}%
\bibitem [{\citenamefont {Heiss}(2012)}]{heiss_physics_2012}%
  \BibitemOpen
  \bibfield  {author} {\bibinfo {author} {\bibfnamefont {W.~D.}\ \bibnamefont
  {Heiss}},\ }\bibfield  {title} {\bibinfo {title} {The physics of exceptional
  points},\ }\href {https://doi.org/10.1088/1751-8113/45/44/444016} {\bibfield
  {journal} {\bibinfo  {journal} {J. Phys. A: Math. Theor.}\ }\textbf {\bibinfo
  {volume} {45}},\ \bibinfo {pages} {444016} (\bibinfo {year}
  {2012})}\BibitemShut {NoStop}%
\bibitem [{\citenamefont {Ashida}\ \emph {et~al.}(2020)\citenamefont {Ashida},
  \citenamefont {Gong},\ and\ \citenamefont
  {Ueda}}]{ashida_non-hermitian_2020}%
  \BibitemOpen
  \bibfield  {author} {\bibinfo {author} {\bibfnamefont {Y.}~\bibnamefont
  {Ashida}}, \bibinfo {author} {\bibfnamefont {Z.}~\bibnamefont {Gong}},\ and\
  \bibinfo {author} {\bibfnamefont {M.}~\bibnamefont {Ueda}},\ }\bibfield
  {title} {\bibinfo {title} {Non-{Hermitian} physics},\ }\href
  {https://doi.org/10.1080/00018732.2021.1876991} {\bibfield  {journal}
  {\bibinfo  {journal} {Adv. Phys.}\ }\textbf {\bibinfo {volume} {69}},\
  \bibinfo {pages} {249} (\bibinfo {year} {2020})}\BibitemShut {NoStop}%
\bibitem [{\citenamefont {Bergholtz}\ \emph {et~al.}(2021)\citenamefont
  {Bergholtz}, \citenamefont {Budich},\ and\ \citenamefont
  {Kunst}}]{bergholtz_exceptional_2021}%
  \BibitemOpen
  \bibfield  {author} {\bibinfo {author} {\bibfnamefont {E.~J.}\ \bibnamefont
  {Bergholtz}}, \bibinfo {author} {\bibfnamefont {J.~C.}\ \bibnamefont
  {Budich}},\ and\ \bibinfo {author} {\bibfnamefont {F.~K.}\ \bibnamefont
  {Kunst}},\ }\bibfield  {title} {\bibinfo {title} {Exceptional topology of
  non-{Hermitian} systems},\ }\href
  {https://doi.org/10.1103/RevModPhys.93.015005} {\bibfield  {journal}
  {\bibinfo  {journal} {Rev. Mod. Phys.}\ }\textbf {\bibinfo {volume} {93}},\
  \bibinfo {pages} {015005} (\bibinfo {year} {2021})}\BibitemShut {NoStop}%
\bibitem [{\citenamefont {Prosen}(2008)}]{prosen_third_2008}%
  \BibitemOpen
  \bibfield  {author} {\bibinfo {author} {\bibfnamefont {T.}~\bibnamefont
  {Prosen}},\ }\bibfield  {title} {\bibinfo {title} {Third quantization: a
  general method to solve master equations for quadratic open {Fermi}
  systems},\ }\href {https://doi.org/10.1088/1367-2630/10/4/043026} {\bibfield
  {journal} {\bibinfo  {journal} {New J. Phys.}\ }\textbf {\bibinfo {volume}
  {10}},\ \bibinfo {pages} {043026} (\bibinfo {year} {2008})}\BibitemShut
  {NoStop}%
\bibitem [{\citenamefont {Prosen}\ and\ \citenamefont
  {{{\v{Z}}}unkovi{\v{c}}}(2010)}]{prosen_exact_2010}%
  \BibitemOpen
  \bibfield  {author} {\bibinfo {author} {\bibfnamefont {T.}~\bibnamefont
  {Prosen}}\ and\ \bibinfo {author} {\bibfnamefont {B.}~\bibnamefont
  {{{\v{Z}}}unkovi{\v{c}}}},\ }\bibfield  {title} {\bibinfo {title} {Exact
  solution of {Markovian} master equations for quadratic {Fermi} systems:
  thermal baths, open {XY} spin chains and non-equilibrium phase transition},\
  }\href {https://doi.org/10.1088/1367-2630/12/2/025016} {\bibfield  {journal}
  {\bibinfo  {journal} {New J. Phys.}\ }\textbf {\bibinfo {volume} {12}},\
  \bibinfo {pages} {025016} (\bibinfo {year} {2010})}\BibitemShut {NoStop}%
\bibitem [{\citenamefont {van Caspel}\ \emph {et~al.}(2019)\citenamefont {van
  Caspel}, \citenamefont {Tapias~Arze},\ and\ \citenamefont
  {P{\'{e}}rez~Castillo}}]{van_caspel_dynamical_2019}%
  \BibitemOpen
  \bibfield  {author} {\bibinfo {author} {\bibfnamefont {M.}~\bibnamefont {van
  Caspel}}, \bibinfo {author} {\bibfnamefont {S.~E.}\ \bibnamefont
  {Tapias~Arze}},\ and\ \bibinfo {author} {\bibfnamefont {I.}~\bibnamefont
  {P{\'{e}}rez~Castillo}},\ }\bibfield  {title} {\bibinfo {title} {Dynamical
  signatures of topological order in the driven-dissipative {Kitaev} chain},\
  }\href {https://doi.org/10.21468/SciPostPhys.6.2.026} {\bibfield  {journal}
  {\bibinfo  {journal} {SciPost Phys.}\ }\textbf {\bibinfo {volume} {6}},\
  \bibinfo {pages} {026} (\bibinfo {year} {2019})}\BibitemShut {NoStop}%
\bibitem [{\citenamefont {Krapivsky}\ \emph {et~al.}(2019)\citenamefont
  {Krapivsky}, \citenamefont {Mallick},\ and\ \citenamefont
  {Sels}}]{krapivsky_free_2019}%
  \BibitemOpen
  \bibfield  {author} {\bibinfo {author} {\bibfnamefont {P.~L.}\ \bibnamefont
  {Krapivsky}}, \bibinfo {author} {\bibfnamefont {K.}~\bibnamefont {Mallick}},\
  and\ \bibinfo {author} {\bibfnamefont {D.}~\bibnamefont {Sels}},\ }\bibfield
  {title} {\bibinfo {title} {Free fermions with a localized source},\ }\href
  {https://doi.org/10.1088/1742-5468/ab4e8e} {\bibfield  {journal} {\bibinfo
  {journal} {J. Stat. Mech.}\ }\textbf {\bibinfo {volume} {2019}},\ \bibinfo
  {pages} {113108} (\bibinfo {year} {2019})}\BibitemShut {NoStop}%
\bibitem [{\citenamefont {Shibata}\ and\ \citenamefont
  {Katsura}(2019)}]{shibata_dissipative_2019}%
  \BibitemOpen
  \bibfield  {author} {\bibinfo {author} {\bibfnamefont {N.}~\bibnamefont
  {Shibata}}\ and\ \bibinfo {author} {\bibfnamefont {H.}~\bibnamefont
  {Katsura}},\ }\bibfield  {title} {\bibinfo {title} {Dissipative quantum
  {Ising} chain as a non-{Hermitian} {Ashkin}-{Teller} model},\ }\href
  {https://doi.org/10.1103/PhysRevB.99.224432} {\bibfield  {journal} {\bibinfo
  {journal} {Phys. Rev. B}\ }\textbf {\bibinfo {volume} {99}},\ \bibinfo
  {pages} {224432} (\bibinfo {year} {2019})}\BibitemShut {NoStop}%
\bibitem [{\citenamefont {Vernier}(2020)}]{vernier_mixing_2020}%
  \BibitemOpen
  \bibfield  {author} {\bibinfo {author} {\bibfnamefont {E.}~\bibnamefont
  {Vernier}},\ }\bibfield  {title} {\bibinfo {title} {Mixing times and cutoffs
  in open quadratic fermionic systems},\ }\href
  {https://doi.org/10.21468/SciPostPhys.9.4.049} {\bibfield  {journal}
  {\bibinfo  {journal} {SciPost Physics}\ }\textbf {\bibinfo {volume} {9}},\
  \bibinfo {pages} {049} (\bibinfo {year} {2020})}\BibitemShut {NoStop}%
\bibitem [{\citenamefont {Lieu}\ \emph {et~al.}(2020)\citenamefont {Lieu},
  \citenamefont {McGinley},\ and\ \citenamefont {Cooper}}]{lieu_tenfold_2020}%
  \BibitemOpen
  \bibfield  {author} {\bibinfo {author} {\bibfnamefont {S.}~\bibnamefont
  {Lieu}}, \bibinfo {author} {\bibfnamefont {M.}~\bibnamefont {McGinley}},\
  and\ \bibinfo {author} {\bibfnamefont {N.~R.}\ \bibnamefont {Cooper}},\
  }\bibfield  {title} {\bibinfo {title} {Tenfold {Way} for {Quadratic}
  {Lindbladians}},\ }\href {https://doi.org/10.1103/PhysRevLett.124.040401}
  {\bibfield  {journal} {\bibinfo  {journal} {Phys. Rev. Lett.}\ }\textbf
  {\bibinfo {volume} {124}},\ \bibinfo {pages} {040401} (\bibinfo {year}
  {2020})}\BibitemShut {NoStop}%
\bibitem [{\citenamefont {Alba}\ and\ \citenamefont
  {Carollo}(2021)}]{alba_noninteracting_2021}%
  \BibitemOpen
  \bibfield  {author} {\bibinfo {author} {\bibfnamefont {V.}~\bibnamefont
  {Alba}}\ and\ \bibinfo {author} {\bibfnamefont {F.}~\bibnamefont {Carollo}},\
  }\bibfield  {title} {\bibinfo {title} {Noninteracting fermionic systems with
  localized dissipation: {Exact} results in the hydrodynamic limit},\ }\href
  {http://arxiv.org/abs/2103.05671} {\bibfield  {journal} {\bibinfo  {journal}
  {arXiv:2103.05671 [cond-mat, physics:quant-ph]}\ } (\bibinfo {year}
  {2021})}\BibitemShut {NoStop}%
\bibitem [{\citenamefont {Prosen}(2011{\natexlab{a}})}]{prosen_open_2011}%
  \BibitemOpen
  \bibfield  {author} {\bibinfo {author} {\bibfnamefont {T.}~\bibnamefont
  {Prosen}},\ }\bibfield  {title} {\bibinfo {title} {Open {XXZ} {Spin} {Chain}:
  {Nonequilibrium} {Steady} {State} and a {Strict} {Bound} on {Ballistic}
  {Transport}},\ }\href {https://doi.org/10.1103/PhysRevLett.106.217206}
  {\bibfield  {journal} {\bibinfo  {journal} {Phys. Rev. Lett.}\ }\textbf
  {\bibinfo {volume} {106}},\ \bibinfo {pages} {217206} (\bibinfo {year}
  {2011}{\natexlab{a}})}\BibitemShut {NoStop}%
\bibitem [{\citenamefont {Prosen}(2011{\natexlab{b}})}]{prosen_exact_2011}%
  \BibitemOpen
  \bibfield  {author} {\bibinfo {author} {\bibfnamefont {T.}~\bibnamefont
  {Prosen}},\ }\bibfield  {title} {\bibinfo {title} {Exact {Nonequilibrium}
  {Steady} {State} of a {Strongly} {Driven} {Open} {XXZ} {Chain}},\ }\href
  {https://doi.org/10.1103/PhysRevLett.107.137201} {\bibfield  {journal}
  {\bibinfo  {journal} {Phys. Rev. Lett.}\ }\textbf {\bibinfo {volume} {107}},\
  \bibinfo {pages} {137201} (\bibinfo {year} {2011}{\natexlab{b}})}\BibitemShut
  {NoStop}%
\bibitem [{\citenamefont {Karevski}\ \emph {et~al.}(2013)\citenamefont
  {Karevski}, \citenamefont {Popkov},\ and\ \citenamefont
  {Sch{\"{u}}tz}}]{karevski_exact_2013}%
  \BibitemOpen
  \bibfield  {author} {\bibinfo {author} {\bibfnamefont {D.}~\bibnamefont
  {Karevski}}, \bibinfo {author} {\bibfnamefont {V.}~\bibnamefont {Popkov}},\
  and\ \bibinfo {author} {\bibfnamefont {G.~M.}\ \bibnamefont {Sch{\"{u}}tz}},\
  }\bibfield  {title} {\bibinfo {title} {Exact {Matrix} {Product} {Solution}
  for the {Boundary}-{Driven} {Lindblad} {XXZ} {Chain}},\ }\href
  {https://doi.org/10.1103/PhysRevLett.110.047201} {\bibfield  {journal}
  {\bibinfo  {journal} {Phys. Rev. Lett.}\ }\textbf {\bibinfo {volume} {110}},\
  \bibinfo {pages} {047201} (\bibinfo {year} {2013})}\BibitemShut {NoStop}%
\bibitem [{\citenamefont {Ilievski}(2017)}]{ilievski_dissipation-driven_2017}%
  \BibitemOpen
  \bibfield  {author} {\bibinfo {author} {\bibfnamefont {E.}~\bibnamefont
  {Ilievski}},\ }\bibfield  {title} {\bibinfo {title} {Dissipation-driven
  integrable fermionic systems: from graded {Yangians} to exact nonequilibrium
  steady states},\ }\href {https://doi.org/10.21468/SciPostPhys.3.4.031}
  {\bibfield  {journal} {\bibinfo  {journal} {SciPost Physics}\ }\textbf
  {\bibinfo {volume} {3}},\ \bibinfo {pages} {031} (\bibinfo {year}
  {2017})}\BibitemShut {NoStop}%
\bibitem [{\citenamefont {Vanicat}\ \emph {et~al.}(2018)\citenamefont
  {Vanicat}, \citenamefont {Zadnik},\ and\ \citenamefont
  {Prosen}}]{vanicat_integrable_2018}%
  \BibitemOpen
  \bibfield  {author} {\bibinfo {author} {\bibfnamefont {M.}~\bibnamefont
  {Vanicat}}, \bibinfo {author} {\bibfnamefont {L.}~\bibnamefont {Zadnik}},\
  and\ \bibinfo {author} {\bibfnamefont {T.}~\bibnamefont {Prosen}},\
  }\bibfield  {title} {\bibinfo {title} {Integrable {Trotterization}: {Local}
  {Conservation} {Laws} and {Boundary} {Driving}},\ }\href
  {https://doi.org/10.1103/PhysRevLett.121.030606} {\bibfield  {journal}
  {\bibinfo  {journal} {Phys. Rev. Lett.}\ }\textbf {\bibinfo {volume} {121}},\
  \bibinfo {pages} {030606} (\bibinfo {year} {2018})}\BibitemShut {NoStop}%
\bibitem [{\citenamefont {Landi}\ \emph {et~al.}(2021)\citenamefont {Landi},
  \citenamefont {Poletti},\ and\ \citenamefont
  {Schaller}}]{landi_non-equilibrium_2021}%
  \BibitemOpen
  \bibfield  {author} {\bibinfo {author} {\bibfnamefont {G.~T.}\ \bibnamefont
  {Landi}}, \bibinfo {author} {\bibfnamefont {D.}~\bibnamefont {Poletti}},\
  and\ \bibinfo {author} {\bibfnamefont {G.}~\bibnamefont {Schaller}},\
  }\bibfield  {title} {\bibinfo {title} {Non-equilibrium boundary driven
  quantum systems: models, methods and properties},\ }\href
  {http://arxiv.org/abs/2104.14350} {\bibfield  {journal} {\bibinfo  {journal}
  {arXiv:2104.14350 [cond-mat, physics:quant-ph]}\ } (\bibinfo {year}
  {2021})}\BibitemShut {NoStop}%
\bibitem [{\citenamefont {Medvedyeva}\ \emph {et~al.}(2016)\citenamefont
  {Medvedyeva}, \citenamefont {Essler},\ and\ \citenamefont
  {Prosen}}]{medvedyeva_exact_2016}%
  \BibitemOpen
  \bibfield  {author} {\bibinfo {author} {\bibfnamefont {M.~V.}\ \bibnamefont
  {Medvedyeva}}, \bibinfo {author} {\bibfnamefont {F.~H.}\ \bibnamefont
  {Essler}},\ and\ \bibinfo {author} {\bibfnamefont {T.}~\bibnamefont
  {Prosen}},\ }\bibfield  {title} {\bibinfo {title} {Exact {Bethe} {Ansatz}
  {Spectrum} of a {Tight}-{Binding} {Chain} with {Dephasing} {Noise}},\ }\href
  {https://doi.org/10.1103/PhysRevLett.117.137202} {\bibfield  {journal}
  {\bibinfo  {journal} {Phys. Rev. Lett.}\ }\textbf {\bibinfo {volume} {117}},\
  \bibinfo {pages} {137202} (\bibinfo {year} {2016})}\BibitemShut {NoStop}%
\bibitem [{\citenamefont {Ziolkowska}\ and\ \citenamefont
  {Essler}(2020)}]{ziolkowska_yang-baxter_2020}%
  \BibitemOpen
  \bibfield  {author} {\bibinfo {author} {\bibfnamefont {A.~A.}\ \bibnamefont
  {Ziolkowska}}\ and\ \bibinfo {author} {\bibfnamefont {F.}~\bibnamefont
  {Essler}},\ }\bibfield  {title} {\bibinfo {title} {Yang-{Baxter} integrable
  {Lindblad} equations},\ }\href {https://doi.org/10.21468/SciPostPhys.8.3.044}
  {\bibfield  {journal} {\bibinfo  {journal} {SciPost Physics}\ }\textbf
  {\bibinfo {volume} {8}},\ \bibinfo {pages} {044} (\bibinfo {year}
  {2020})}\BibitemShut {NoStop}%
\bibitem [{\citenamefont {Bu{\v{c}}a}\ \emph {et~al.}(2020)\citenamefont
  {Bu{\v{c}}a}, \citenamefont {Booker}, \citenamefont {Medenjak},\ and\
  \citenamefont {Jaksch}}]{buca_bethe_2020}%
  \BibitemOpen
  \bibfield  {author} {\bibinfo {author} {\bibfnamefont {B.}~\bibnamefont
  {Bu{\v{c}}a}}, \bibinfo {author} {\bibfnamefont {C.}~\bibnamefont {Booker}},
  \bibinfo {author} {\bibfnamefont {M.}~\bibnamefont {Medenjak}},\ and\
  \bibinfo {author} {\bibfnamefont {D.}~\bibnamefont {Jaksch}},\ }\bibfield
  {title} {\bibinfo {title} {Bethe ansatz approach for dissipation: exact
  solutions of quantum many-body dynamics under loss},\ }\href
  {https://doi.org/10.1088/1367-2630/abd124} {\bibfield  {journal} {\bibinfo
  {journal} {New J. Phys.}\ }\textbf {\bibinfo {volume} {22}},\ \bibinfo
  {pages} {123040} (\bibinfo {year} {2020})}\BibitemShut {NoStop}%
\bibitem [{\citenamefont {de~Leeuw}\ \emph {et~al.}(2021)\citenamefont
  {de~Leeuw}, \citenamefont {Paletta},\ and\ \citenamefont
  {Pozsgay}}]{de_leeuw_constructing_2021}%
  \BibitemOpen
  \bibfield  {author} {\bibinfo {author} {\bibfnamefont {M.}~\bibnamefont
  {de~Leeuw}}, \bibinfo {author} {\bibfnamefont {C.}~\bibnamefont {Paletta}},\
  and\ \bibinfo {author} {\bibfnamefont {B.}~\bibnamefont {Pozsgay}},\
  }\bibfield  {title} {\bibinfo {title} {Constructing {Integrable} {Lindblad}
  {Superoperators}},\ }\href {https://doi.org/10.1103/PhysRevLett.126.240403}
  {\bibfield  {journal} {\bibinfo  {journal} {Phys. Rev. Lett.}\ }\textbf
  {\bibinfo {volume} {126}},\ \bibinfo {pages} {240403} (\bibinfo {year}
  {2021})}\BibitemShut {NoStop}%
\bibitem [{\citenamefont {Rowlands}\ and\ \citenamefont
  {Lamacraft}(2018)}]{rowlands_noisy_2018}%
  \BibitemOpen
  \bibfield  {author} {\bibinfo {author} {\bibfnamefont {D.~A.}\ \bibnamefont
  {Rowlands}}\ and\ \bibinfo {author} {\bibfnamefont {A.}~\bibnamefont
  {Lamacraft}},\ }\bibfield  {title} {\bibinfo {title} {Noisy {Spins} and the
  {Richardson}-{Gaudin} {Model}},\ }\href
  {https://doi.org/10.1103/PhysRevLett.120.090401} {\bibfield  {journal}
  {\bibinfo  {journal} {Phys. Rev. Lett.}\ }\textbf {\bibinfo {volume} {120}},\
  \bibinfo {pages} {090401} (\bibinfo {year} {2018})}\BibitemShut {NoStop}%
\bibitem [{\citenamefont {Ribeiro}\ and\ \citenamefont
  {Prosen}(2019)}]{ribeiro_integrable_2019}%
  \BibitemOpen
  \bibfield  {author} {\bibinfo {author} {\bibfnamefont {P.}~\bibnamefont
  {Ribeiro}}\ and\ \bibinfo {author} {\bibfnamefont {T.}~\bibnamefont
  {Prosen}},\ }\bibfield  {title} {\bibinfo {title} {Integrable {Quantum}
  {Dynamics} of {Open} {Collective} {Spin} {Models}},\ }\href
  {https://doi.org/10.1103/PhysRevLett.122.010401} {\bibfield  {journal}
  {\bibinfo  {journal} {Phys. Rev. Lett.}\ }\textbf {\bibinfo {volume} {122}},\
  \bibinfo {pages} {010401} (\bibinfo {year} {2019})}\BibitemShut {NoStop}%
\bibitem [{\citenamefont {Lerma-Hern{\'{a}}ndez}\ \emph
  {et~al.}(2020)\citenamefont {Lerma-Hern{\'{a}}ndez}, \citenamefont
  {Rubio-Garc{\'{i}}a},\ and\ \citenamefont
  {Dukelsky}}]{lerma-hernandez_trigonometric_2020}%
  \BibitemOpen
  \bibfield  {author} {\bibinfo {author} {\bibfnamefont {S.}~\bibnamefont
  {Lerma-Hern{\'{a}}ndez}}, \bibinfo {author} {\bibfnamefont {A.}~\bibnamefont
  {Rubio-Garc{\'{i}}a}},\ and\ \bibinfo {author} {\bibfnamefont
  {J.}~\bibnamefont {Dukelsky}},\ }\bibfield  {title} {\bibinfo {title}
  {Trigonometric {SU}({N}) {Richardson}-{Gaudin} models and dissipative
  multi-level atomic systems},\ }\href
  {https://doi.org/10.1088/1751-8121/abab54} {\bibfield  {journal} {\bibinfo
  {journal} {J. Phys. A: Math. Theor.}\ }\textbf {\bibinfo {volume} {53}},\
  \bibinfo {pages} {395302} (\bibinfo {year} {2020})}\BibitemShut {NoStop}%
\bibitem [{\citenamefont {Rubio-Garc{\'{i}}a}\ \emph
  {et~al.}(2021)\citenamefont {Rubio-Garc{\'{i}}a}, \citenamefont {Molina},\
  and\ \citenamefont {Dukelsky}}]{rubio-garcia_integrability_2021}%
  \BibitemOpen
  \bibfield  {author} {\bibinfo {author} {\bibfnamefont {{\'{A}}.}~\bibnamefont
  {Rubio-Garc{\'{i}}a}}, \bibinfo {author} {\bibfnamefont {R.~A.}\ \bibnamefont
  {Molina}},\ and\ \bibinfo {author} {\bibfnamefont {J.}~\bibnamefont
  {Dukelsky}},\ }\bibfield  {title} {\bibinfo {title} {From integrability to
  chaos in quantum {Liouvillians}},\ }\href {http://arxiv.org/abs/2102.13452}
  {\bibfield  {journal} {\bibinfo  {journal} {arXiv:2102.13452 [cond-mat,
  physics:nlin, physics:quant-ph]}\ } (\bibinfo {year} {2021})}\BibitemShut
  {NoStop}%
\bibitem [{\citenamefont {Ortiz}\ \emph {et~al.}(2005)\citenamefont {Ortiz},
  \citenamefont {Somma}, \citenamefont {Dukelsky},\ and\ \citenamefont
  {Rombouts}}]{ortiz_exactly-solvable_2005}%
  \BibitemOpen
  \bibfield  {author} {\bibinfo {author} {\bibfnamefont {G.}~\bibnamefont
  {Ortiz}}, \bibinfo {author} {\bibfnamefont {R.}~\bibnamefont {Somma}},
  \bibinfo {author} {\bibfnamefont {J.}~\bibnamefont {Dukelsky}},\ and\
  \bibinfo {author} {\bibfnamefont {S.}~\bibnamefont {Rombouts}},\ }\bibfield
  {title} {\bibinfo {title} {Exactly-solvable models derived from a generalized
  {Gaudin} algebra},\ }\href {https://doi.org/10.1016/j.nuclphysb.2004.11.008}
  {\bibfield  {journal} {\bibinfo  {journal} {Nucl. Phys. B}\ }\textbf
  {\bibinfo {volume} {707}},\ \bibinfo {pages} {421} (\bibinfo {year}
  {2005})}\BibitemShut {NoStop}%
\bibitem [{\citenamefont {Iba{\~n}ez}\ \emph {et~al.}(2009)\citenamefont
  {Iba{\~n}ez}, \citenamefont {Links}, \citenamefont {Sierra},\ and\
  \citenamefont {Zhao}}]{ibanez_exactly_2009}%
  \BibitemOpen
  \bibfield  {author} {\bibinfo {author} {\bibfnamefont {M.}~\bibnamefont
  {Iba{\~n}ez}}, \bibinfo {author} {\bibfnamefont {J.}~\bibnamefont {Links}},
  \bibinfo {author} {\bibfnamefont {G.}~\bibnamefont {Sierra}},\ and\ \bibinfo
  {author} {\bibfnamefont {S.-Y.}\ \bibnamefont {Zhao}},\ }\bibfield  {title}
  {\bibinfo {title} {{E}xactly solvable pairing model for superconductors with
  $p_x+ip_y$-wave symmetry},\ }\href
  {https://doi.org/10.1103/PhysRevB.79.180501} {\bibfield  {journal} {\bibinfo
  {journal} {Phys. Rev. B}\ }\textbf {\bibinfo {volume} {79}},\ \bibinfo
  {pages} {180501} (\bibinfo {year} {2009})}\BibitemShut {NoStop}%
\bibitem [{\citenamefont {Rombouts}\ \emph {et~al.}(2010)\citenamefont
  {Rombouts}, \citenamefont {Dukelsky},\ and\ \citenamefont
  {Ortiz}}]{rombouts_quantum_2010}%
  \BibitemOpen
  \bibfield  {author} {\bibinfo {author} {\bibfnamefont {S.~M.~A.}\
  \bibnamefont {Rombouts}}, \bibinfo {author} {\bibfnamefont {J.}~\bibnamefont
  {Dukelsky}},\ and\ \bibinfo {author} {\bibfnamefont {G.}~\bibnamefont
  {Ortiz}},\ }\bibfield  {title} {\bibinfo {title} {{Q}uantum phase diagram of
  the integrable $p_x+ip_y$ fermionic superfluid},\ }\href
  {https://doi.org/10.1103/PhysRevB.82.224510} {\bibfield  {journal} {\bibinfo
  {journal} {Phys. Rev. B}\ }\textbf {\bibinfo {volume} {82}},\ \bibinfo
  {pages} {224510} (\bibinfo {year} {2010})}\BibitemShut {NoStop}%
\bibitem [{\citenamefont {Van~Raemdonck}\ \emph {et~al.}(2014)\citenamefont
  {Van~Raemdonck}, \citenamefont {De~Baerdemacker},\ and\ \citenamefont
  {Van~Neck}}]{van_raemdonck_exact_2014}%
  \BibitemOpen
  \bibfield  {author} {\bibinfo {author} {\bibfnamefont {M.}~\bibnamefont
  {Van~Raemdonck}}, \bibinfo {author} {\bibfnamefont {S.}~\bibnamefont
  {De~Baerdemacker}},\ and\ \bibinfo {author} {\bibfnamefont {D.}~\bibnamefont
  {Van~Neck}},\ }\bibfield  {title} {\bibinfo {title} {{E}xact solution of the
  $p_x+ip_y$ pairing {Hamiltonian} by deforming the pairing algebra},\ }\href
  {https://doi.org/10.1103/PhysRevB.89.155136} {\bibfield  {journal} {\bibinfo
  {journal} {Phys. Rev. B}\ }\textbf {\bibinfo {volume} {89}},\ \bibinfo
  {pages} {155136} (\bibinfo {year} {2014})}\BibitemShut {NoStop}%
\bibitem [{\citenamefont {Links}\ \emph {et~al.}(2015)\citenamefont {Links},
  \citenamefont {Marquette},\ and\ \citenamefont
  {Moghaddam}}]{links_exact_2015}%
  \BibitemOpen
  \bibfield  {author} {\bibinfo {author} {\bibfnamefont {J.}~\bibnamefont
  {Links}}, \bibinfo {author} {\bibfnamefont {I.}~\bibnamefont {Marquette}},\
  and\ \bibinfo {author} {\bibfnamefont {A.}~\bibnamefont {Moghaddam}},\
  }\bibfield  {title} {\bibinfo {title} {{E}xact solution of the $p+ip$
  {Hamiltonian} revisited: duality relations in the hole-pair picture},\ }\href
  {https://doi.org/10.1088/1751-8113/48/37/374001} {\bibfield  {journal}
  {\bibinfo  {journal} {J. Phys. A: Math. Theor.}\ }\textbf {\bibinfo {volume}
  {48}},\ \bibinfo {pages} {374001} (\bibinfo {year} {2015})}\BibitemShut
  {NoStop}%
\bibitem [{\citenamefont {Claeys}\ \emph {et~al.}(2016)\citenamefont {Claeys},
  \citenamefont {De~Baerdemacker},\ and\ \citenamefont
  {Van~Neck}}]{claeys_read-green_2016}%
  \BibitemOpen
  \bibfield  {author} {\bibinfo {author} {\bibfnamefont {P.~W.}\ \bibnamefont
  {Claeys}}, \bibinfo {author} {\bibfnamefont {S.}~\bibnamefont
  {De~Baerdemacker}},\ and\ \bibinfo {author} {\bibfnamefont {D.}~\bibnamefont
  {Van~Neck}},\ }\bibfield  {title} {\bibinfo {title} {{R}ead-{Green}
  resonances in a topological superconductor coupled to a bath},\ }\href
  {https://doi.org/10.1103/PhysRevB.93.220503} {\bibfield  {journal} {\bibinfo
  {journal} {Phys. Rev. B}\ }\textbf {\bibinfo {volume} {93}},\ \bibinfo
  {pages} {220503} (\bibinfo {year} {2016})}\BibitemShut {NoStop}%
\bibitem [{\citenamefont
  {Mostafazadeh}(2002)}]{mostafazadeh_pseudo-hermiticity_2002}%
  \BibitemOpen
  \bibfield  {author} {\bibinfo {author} {\bibfnamefont {A.}~\bibnamefont
  {Mostafazadeh}},\ }\bibfield  {title} {\bibinfo {title} {Pseudo-{Hermiticity}
  versus {PT} symmetry: {The} necessary condition for the reality of the
  spectrum of a non-{Hermitian} {Hamiltonian}},\ }\href
  {https://doi.org/10.1063/1.1418246} {\bibfield  {journal} {\bibinfo
  {journal} {J. Math. Phys.}\ }\textbf {\bibinfo {volume} {43}},\ \bibinfo
  {pages} {205} (\bibinfo {year} {2002})}\BibitemShut {NoStop}%
\bibitem [{Note1()}]{Note1}%
  \BibitemOpen
  \bibinfo {note} {Technically, this operator generates the evolution of the
  corresponding correlation functions, and its Hermitian conjugate generates
  the evolution of the operators. We write the Liouvillian in this way in order
  to make the connection with the literature.}\BibitemShut {Stop}%
\bibitem [{\citenamefont {Bender}\ and\ \citenamefont
  {Boettcher}(1998)}]{bender_real_1998}%
  \BibitemOpen
  \bibfield  {author} {\bibinfo {author} {\bibfnamefont {C.~M.}\ \bibnamefont
  {Bender}}\ and\ \bibinfo {author} {\bibfnamefont {S.}~\bibnamefont
  {Boettcher}},\ }\bibfield  {title} {\bibinfo {title} {Real {Spectra} in
  {Non}-{Hermitian} {Hamiltonians} {Having} {P} {T} {Symmetry}},\ }\href
  {https://doi.org/10.1103/PhysRevLett.80.5243} {\bibfield  {journal} {\bibinfo
   {journal} {Phys. Rev. Lett.}\ }\textbf {\bibinfo {volume} {80}},\ \bibinfo
  {pages} {5243} (\bibinfo {year} {1998})}\BibitemShut {NoStop}%
\bibitem [{\citenamefont {Prosen}(2012)}]{prosen_p_2012}%
  \BibitemOpen
  \bibfield  {author} {\bibinfo {author} {\bibfnamefont {T.}~\bibnamefont
  {Prosen}},\ }\bibfield  {title} {\bibinfo {title} {{PT} -{Symmetric}
  {Quantum} {Liouvillean} {Dynamics}},\ }\href
  {https://doi.org/10.1103/PhysRevLett.109.090404} {\bibfield  {journal}
  {\bibinfo  {journal} {Phys. Rev. Lett.}\ }\textbf {\bibinfo {volume} {109}},\
  \bibinfo {pages} {090404} (\bibinfo {year} {2012})}\BibitemShut {NoStop}%
\bibitem [{\citenamefont {Hirsch}\ \emph {et~al.}(2002)\citenamefont {Hirsch},
  \citenamefont {Mariano}, \citenamefont {Dukelsky},\ and\ \citenamefont
  {Schuck}}]{hirsch_fully_2002}%
  \BibitemOpen
  \bibfield  {author} {\bibinfo {author} {\bibfnamefont {J.~G.}\ \bibnamefont
  {Hirsch}}, \bibinfo {author} {\bibfnamefont {A.}~\bibnamefont {Mariano}},
  \bibinfo {author} {\bibfnamefont {J.}~\bibnamefont {Dukelsky}},\ and\
  \bibinfo {author} {\bibfnamefont {P.}~\bibnamefont {Schuck}},\ }\bibfield
  {title} {\bibinfo {title} {Fully {Self}-{Consistent} {RPA} {Description} of
  the {Many} {Level} {Pairing} {Model}},\ }\href
  {https://doi.org/10.1006/aphy.2002.6230} {\bibfield  {journal} {\bibinfo
  {journal} {Ann. Phys.}\ }\textbf {\bibinfo {volume} {296}},\ \bibinfo {pages}
  {187} (\bibinfo {year} {2002})}\BibitemShut {NoStop}%
\bibitem [{\citenamefont {Andrews}\ and\ \citenamefont
  {Thirunamachandran}(1977)}]{andrews_threedimensional_1977}%
  \BibitemOpen
  \bibfield  {author} {\bibinfo {author} {\bibfnamefont {D.~L.}\ \bibnamefont
  {Andrews}}\ and\ \bibinfo {author} {\bibfnamefont {T.}~\bibnamefont
  {Thirunamachandran}},\ }\bibfield  {title} {\bibinfo {title} {On
  three-dimensional rotational averages},\ }\href
  {https://doi.org/10.1063/1.434725} {\bibfield  {journal} {\bibinfo  {journal}
  {J. Chem. Phys.}\ }\textbf {\bibinfo {volume} {67}},\ \bibinfo {pages} {5026}
  (\bibinfo {year} {1977})}\BibitemShut {NoStop}%
\bibitem [{\citenamefont {Bernhart}(1999)}]{bernhart_catalan_1999}%
  \BibitemOpen
  \bibfield  {author} {\bibinfo {author} {\bibfnamefont {F.~R.}\ \bibnamefont
  {Bernhart}},\ }\bibfield  {title} {\bibinfo {title} {Catalan, {Motzkin}, and
  {Riordan} numbers},\ }\href {https://doi.org/10.1016/S0012-365X(99)00054-0}
  {\bibfield  {journal} {\bibinfo  {journal} {Discrete Mathematics}\ }\bibinfo
  {series} {Selected papers in honor of {Henry} {W}. {Gould}},\ \textbf
  {\bibinfo {volume} {204}},\ \bibinfo {pages} {73} (\bibinfo {year}
  {1999})}\BibitemShut {NoStop}%
\bibitem [{\citenamefont
  {De~Baerdemacker}(2012)}]{de_baerdemacker_richardson-gaudin_2012}%
  \BibitemOpen
  \bibfield  {author} {\bibinfo {author} {\bibfnamefont {S.}~\bibnamefont
  {De~Baerdemacker}},\ }\bibfield  {title} {\bibinfo {title}
  {Richardson-{Gaudin} integrability in the contraction limit of the
  quasispin},\ }\href {https://doi.org/10.1103/PhysRevC.86.044332} {\bibfield
  {journal} {\bibinfo  {journal} {Phys. Rev. C}\ }\textbf {\bibinfo {volume}
  {86}},\ \bibinfo {pages} {044332} (\bibinfo {year} {2012})}\BibitemShut
  {NoStop}%
\bibitem [{\citenamefont {Claeys}\ \emph {et~al.}(2015)\citenamefont {Claeys},
  \citenamefont {De~Baerdemacker}, \citenamefont {Van~Raemdonck},\ and\
  \citenamefont {Van~Neck}}]{claeys_eigenvalue-based_2015}%
  \BibitemOpen
  \bibfield  {author} {\bibinfo {author} {\bibfnamefont {P.~W.}\ \bibnamefont
  {Claeys}}, \bibinfo {author} {\bibfnamefont {S.}~\bibnamefont
  {De~Baerdemacker}}, \bibinfo {author} {\bibfnamefont {M.}~\bibnamefont
  {Van~Raemdonck}},\ and\ \bibinfo {author} {\bibfnamefont {D.}~\bibnamefont
  {Van~Neck}},\ }\bibfield  {title} {\bibinfo {title} {{E}igenvalue-based
  method and form-factor determinant representations for integrable {XXZ}
  {Richardson}-{Gaudin} models},\ }\href
  {https://doi.org/10.1103/PhysRevB.91.155102} {\bibfield  {journal} {\bibinfo
  {journal} {Phys. Rev. B}\ }\textbf {\bibinfo {volume} {91}},\ \bibinfo
  {pages} {155102} (\bibinfo {year} {2015})}\BibitemShut {NoStop}%
\bibitem [{\citenamefont {Babelon}\ and\ \citenamefont
  {Talalaev}(2007)}]{babelon_bethe_2007}%
  \BibitemOpen
  \bibfield  {author} {\bibinfo {author} {\bibfnamefont {O.}~\bibnamefont
  {Babelon}}\ and\ \bibinfo {author} {\bibfnamefont {D.}~\bibnamefont
  {Talalaev}},\ }\bibfield  {title} {\bibinfo {title} {On the {Bethe} ansatz
  for the {Jaynes}-{Cummings}-{Gaudin} model},\ }\href
  {https://doi.org/10.1088/1742-5468/2007/06/P06013} {\bibfield  {journal}
  {\bibinfo  {journal} {J. Stat. Mech.}\ }\textbf {\bibinfo {volume} {2007}},\
  \bibinfo {pages} {P06013} (\bibinfo {year} {2007})}\BibitemShut {NoStop}%
\bibitem [{\citenamefont {Faribault}\ \emph {et~al.}(2011)\citenamefont
  {Faribault}, \citenamefont {El~Araby}, \citenamefont {Str{\"a}ter},\ and\
  \citenamefont {Gritsev}}]{faribault_gaudin_2011}%
  \BibitemOpen
  \bibfield  {author} {\bibinfo {author} {\bibfnamefont {A.}~\bibnamefont
  {Faribault}}, \bibinfo {author} {\bibfnamefont {O.}~\bibnamefont {El~Araby}},
  \bibinfo {author} {\bibfnamefont {C.}~\bibnamefont {Str{\"a}ter}},\ and\
  \bibinfo {author} {\bibfnamefont {V.}~\bibnamefont {Gritsev}},\ }\bibfield
  {title} {\bibinfo {title} {{G}audin models solver based on the correspondence
  between {Bethe} ansatz and ordinary differential equations},\ }\href
  {https://doi.org/10.1103/PhysRevB.83.235124} {\bibfield  {journal} {\bibinfo
  {journal} {Phys. Rev. B}\ }\textbf {\bibinfo {volume} {83}},\ \bibinfo
  {pages} {235124} (\bibinfo {year} {2011})}\BibitemShut {NoStop}%
\bibitem [{\citenamefont {El~Araby}\ \emph {et~al.}(2012)\citenamefont
  {El~Araby}, \citenamefont {Gritsev},\ and\ \citenamefont
  {Faribault}}]{el_araby_bethe_2012}%
  \BibitemOpen
  \bibfield  {author} {\bibinfo {author} {\bibfnamefont {O.}~\bibnamefont
  {El~Araby}}, \bibinfo {author} {\bibfnamefont {V.}~\bibnamefont {Gritsev}},\
  and\ \bibinfo {author} {\bibfnamefont {A.}~\bibnamefont {Faribault}},\
  }\bibfield  {title} {\bibinfo {title} {{B}ethe ansatz and ordinary
  differential equation correspondence for degenerate {Gaudin} models},\ }\href
  {https://doi.org/10.1103/PhysRevB.85.115130} {\bibfield  {journal} {\bibinfo
  {journal} {Phys. Rev. B}\ }\textbf {\bibinfo {volume} {85}},\ \bibinfo
  {pages} {115130} (\bibinfo {year} {2012})}\BibitemShut {NoStop}%
\bibitem [{\citenamefont {Claeys}\ \emph {et~al.}(2017)\citenamefont {Claeys},
  \citenamefont {De~Baerdemacker},\ and\ \citenamefont
  {Van~Neck}}]{claeys_inner_2017}%
  \BibitemOpen
  \bibfield  {author} {\bibinfo {author} {\bibfnamefont {P.~W.}\ \bibnamefont
  {Claeys}}, \bibinfo {author} {\bibfnamefont {S.}~\bibnamefont
  {De~Baerdemacker}},\ and\ \bibinfo {author} {\bibfnamefont {D.}~\bibnamefont
  {Van~Neck}},\ }\bibfield  {title} {\bibinfo {title} {{I}nner products in
  integrable {Richardson-Gaudin} models},\ }\href
  {https://doi.org/10.21468/SciPostPhys.3.4.028} {\bibfield  {journal}
  {\bibinfo  {journal} {SciPost Phys.}\ }\textbf {\bibinfo {volume} {3}},\
  \bibinfo {pages} {028} (\bibinfo {year} {2017})}\BibitemShut {NoStop}%
\bibitem [{\citenamefont {Claeys}(2018)}]{claeys_richardson-gaudin_2018}%
  \BibitemOpen
  \bibfield  {author} {\bibinfo {author} {\bibfnamefont {P.~W.}\ \bibnamefont
  {Claeys}},\ }\emph {\bibinfo {title} {Richardson-{Gaudin} models and broken
  integrability}},\ \href {http://arxiv.org/abs/1809.04447} {Ph.D. thesis},\
  \bibinfo  {school} {Ghent University} (\bibinfo {year} {2018}),\ \bibinfo
  {note} {arXiv: 1809.04447}\BibitemShut {NoStop}%
\bibitem [{\citenamefont {Dimo}\ and\ \citenamefont
  {Faribault}(2018)}]{dimo_quadratic_2018}%
  \BibitemOpen
  \bibfield  {author} {\bibinfo {author} {\bibfnamefont {C.}~\bibnamefont
  {Dimo}}\ and\ \bibinfo {author} {\bibfnamefont {A.}~\bibnamefont
  {Faribault}},\ }\bibfield  {title} {\bibinfo {title} {Quadratic operator
  relations and {Bethe} equations for spin-1/2 {Richardson}-{Gaudin} models},\
  }\href {https://doi.org/10.1088/1751-8121/aaccb4} {\bibfield  {journal}
  {\bibinfo  {journal} {J. Phys. A: Math. Theor.}\ }\textbf {\bibinfo {volume}
  {51}},\ \bibinfo {pages} {325202} (\bibinfo {year} {2018})}\BibitemShut
  {NoStop}%
\bibitem [{\citenamefont {Faribault}\ and\ \citenamefont
  {Dimo}(2018)}]{faribault_bethe-ansatz-free_2018}%
  \BibitemOpen
  \bibfield  {author} {\bibinfo {author} {\bibfnamefont {A.}~\bibnamefont
  {Faribault}}\ and\ \bibinfo {author} {\bibfnamefont {C.}~\bibnamefont
  {Dimo}},\ }\bibfield  {title} {\bibinfo {title} {``{Bethe}-{Ansatz}-free''
  eigenstates of spin-1/2 {Richardson}-{Gaudin} integrable models},\ }\href
  {http://arxiv.org/abs/1812.06428} {\bibfield  {journal} {\bibinfo  {journal}
  {arXiv:1812.06428 [math-ph]}\ } (\bibinfo {year} {2018})}\BibitemShut
  {NoStop}%
\bibitem [{\citenamefont {Claeys}\ \emph {et~al.}(2019)\citenamefont {Claeys},
  \citenamefont {Dimo}, \citenamefont {Baerdemacker},\ and\ \citenamefont
  {Faribault}}]{claeys_integrable_2019}%
  \BibitemOpen
  \bibfield  {author} {\bibinfo {author} {\bibfnamefont {P.~W.}\ \bibnamefont
  {Claeys}}, \bibinfo {author} {\bibfnamefont {C.}~\bibnamefont {Dimo}},
  \bibinfo {author} {\bibfnamefont {S.~D.}\ \bibnamefont {Baerdemacker}},\ and\
  \bibinfo {author} {\bibfnamefont {A.}~\bibnamefont {Faribault}},\ }\bibfield
  {title} {\bibinfo {title} {Integrable spin-1/2 {Richardson}-{Gaudin} {XYZ}
  models in an arbitrary magnetic field},\ }\href
  {https://doi.org/10.1088/1751-8121/aafe9b} {\bibfield  {journal} {\bibinfo
  {journal} {J. Phys. A: Math. Theor.}\ }\textbf {\bibinfo {volume} {52}},\
  \bibinfo {pages} {08LT01} (\bibinfo {year} {2019})}\BibitemShut {NoStop}%
\bibitem [{\citenamefont {D'Alessio}\ \emph {et~al.}(2016)\citenamefont
  {D'Alessio}, \citenamefont {Kafri}, \citenamefont {Polkovnikov},\ and\
  \citenamefont {Rigol}}]{dalessio_quantum_2016}%
  \BibitemOpen
  \bibfield  {author} {\bibinfo {author} {\bibfnamefont {L.}~\bibnamefont
  {D'Alessio}}, \bibinfo {author} {\bibfnamefont {Y.}~\bibnamefont {Kafri}},
  \bibinfo {author} {\bibfnamefont {A.}~\bibnamefont {Polkovnikov}},\ and\
  \bibinfo {author} {\bibfnamefont {M.}~\bibnamefont {Rigol}},\ }\bibfield
  {title} {\bibinfo {title} {From quantum chaos and eigenstate thermalization
  to statistical mechanics and thermodynamics},\ }\href
  {https://doi.org/10.1080/00018732.2016.1198134} {\bibfield  {journal}
  {\bibinfo  {journal} {Adv. Phys.}\ }\textbf {\bibinfo {volume} {65}},\
  \bibinfo {pages} {239} (\bibinfo {year} {2016})}\BibitemShut {NoStop}%
\bibitem [{\citenamefont {Villazon}\ \emph {et~al.}(2020)\citenamefont
  {Villazon}, \citenamefont {Chandran},\ and\ \citenamefont
  {Claeys}}]{villazon_integrability_2020}%
  \BibitemOpen
  \bibfield  {author} {\bibinfo {author} {\bibfnamefont {T.}~\bibnamefont
  {Villazon}}, \bibinfo {author} {\bibfnamefont {A.}~\bibnamefont {Chandran}},\
  and\ \bibinfo {author} {\bibfnamefont {P.~W.}\ \bibnamefont {Claeys}},\
  }\bibfield  {title} {\bibinfo {title} {Integrability and dark states in an
  anisotropic central spin model},\ }\href
  {https://doi.org/10.1103/PhysRevResearch.2.032052} {\bibfield  {journal}
  {\bibinfo  {journal} {Phys. Rev. Research}\ }\textbf {\bibinfo {volume}
  {2}},\ \bibinfo {pages} {032052} (\bibinfo {year} {2020})}\BibitemShut
  {NoStop}%
\bibitem [{\citenamefont {Stieltjes}(1885)}]{stieltjes_theoreme_1885}%
  \BibitemOpen
  \bibfield  {author} {\bibinfo {author} {\bibfnamefont {T.~J.}\ \bibnamefont
  {Stieltjes}},\ }\bibfield  {title} {\bibinfo {title} {Un th{\'{e}}or{\`{e}}me
  d'alg{\`{e}}bre},\ }\href {https://doi.org/10.1007/BF02400420} {\bibfield
  {journal} {\bibinfo  {journal} {Acta Math.}\ }\textbf {\bibinfo {volume}
  {6}},\ \bibinfo {pages} {319} (\bibinfo {year} {1885})}\BibitemShut {NoStop}%
\bibitem [{\citenamefont {Sriram~Shastry}\ and\ \citenamefont
  {Dhar}(2001)}]{sriram_shastry_solution_2001}%
  \BibitemOpen
  \bibfield  {author} {\bibinfo {author} {\bibfnamefont {B.}~\bibnamefont
  {Sriram~Shastry}}\ and\ \bibinfo {author} {\bibfnamefont {A.}~\bibnamefont
  {Dhar}},\ }\bibfield  {title} {\bibinfo {title} {Solution of a generalized
  {Stieltjes} problem},\ }\href {https://doi.org/10.1088/0305-4470/34/31/313}
  {\bibfield  {journal} {\bibinfo  {journal} {J. Phys. A}\ }\textbf {\bibinfo
  {volume} {34}},\ \bibinfo {pages} {6197} (\bibinfo {year}
  {2001})}\BibitemShut {NoStop}%
\end{thebibliography}%
\end{document}